\documentclass[journal]{IEEEtran}

\usepackage[noadjust]{cite}

\ifCLASSINFOpdf
   \usepackage[pdftex]{graphicx}
   \DeclareGraphicsExtensions{.pdf,.jpeg,.png}
\else
   \usepackage[dvips]{graphicx}
   \DeclareGraphicsExtensions{.eps}
\fi

\usepackage[cmex10]{amsmath}
\usepackage{amssymb,bm,url,subcaption}
\begin{document}
%
\title{Principles of Physical Layer Security in Multiuser Wireless Networks: A Survey}

\author{Amitav~Mukherjee, \emph{Member,~IEEE,} S. Ali A. Fakoorian, \emph{Student~Member,~IEEE,} Jing~Huang, \emph{Member,~IEEE,} and A.~Lee~Swindlehurst, \emph{Fellow,~IEEE}
\thanks{A. Mukherjee is with the Wireless Systems Research Lab of Hitachi America, Ltd., Santa Clara, CA 95050, USA (e-mail: \tt{amitav.mukherjee@hal.hitachi.com}).}
\thanks{S. A. A. Fakoorian is with Qualcomm Corporate R\&D, San Diego, CA 92121, USA  (email: {\tt afakoori@uci.edu}).}
\thanks{J. Huang is with Qualcomm Technologies Inc., Santa Clara, CA 95051, USA (e-mail: {\tt jinghuang@qti.qualcomm.com}).}
\thanks{A. L. Swindlehurst is with the Center for Pervasive Communications and Computing, University of California, Irvine, CA 92697-2625, USA (e-mail:  {\tt swindle@uci.edu}).}
\thanks{This work was supported by the National Science Foundation under grant CCF-1117983.}
}
\maketitle

\begin{abstract}
This paper provides a comprehensive review of the domain of physical layer security in multiuser wireless networks. The essential premise of physical layer security is to enable the exchange of confidential messages over a wireless medium in the presence of unauthorized eavesdroppers, without relying on higher-layer encryption. This can be achieved primarily in two ways: without the need for a secret key by intelligently designing transmit coding strategies, or by exploiting the wireless communication medium to develop secret keys over public channels. The survey begins with an overview of the foundations dating back to the pioneering work of Shannon and Wyner on information-theoretic security. We then describe the evolution of secure transmission strategies from point-to-point channels to multiple-antenna systems, followed by generalizations to multiuser broadcast, multiple-access, interference, and relay networks. Secret-key generation and establishment protocols based on physical layer mechanisms are subsequently covered. Approaches for secrecy based on channel coding design are then examined, along with a description of inter-disciplinary approaches based on game theory and stochastic geometry. The associated problem of physical layer message authentication is also briefly introduced. The survey concludes with observations on potential research directions in this area.
\end{abstract}


\section{INTRODUCTION}
The two fundamental characteristics of the wireless medium, namely \emph{broadcast} and \emph{superposition}, present different challenges in ensuring reliable and/or secure communications in the presence of adversarial users. The broadcast nature of wireless communications makes it difficult to shield transmitted signals from unintended recipients, while superposition can lead to the overlapping of multiple signals at the receiver.
As a consequence, adversarial users are commonly modeled either as (1) an unauthorized receiver that tries to extract information from an ongoing transmission without being detected, or (2) a malicious transmitter (\emph{jammer}) that tries to degrade the signal at the intended receiver \cite{Stark88}-\cite{Basar04}.

While jamming and counter-jamming physical layer strategies have been of long-standing interest especially in military networks, the security of data transmission has traditionally been entrusted to key-based enciphering (cryptographic) techniques at the network layer \cite{Massey88}. However, in dynamic wireless networks
this raises issues such as key distribution for symmetric
cryptosystems, and high computational complexity of asymmetric
cryptosystems. More importantly, all cryptographic measures are based
on the premise that it is computationally infeasible for them to be
deciphered without knowledge of the secret key, which remains
mathematically unproven. Ciphers that were considered virtually unbreakable in the past are continually surmounted due to the relentless growth of computational power. Thus, the vulnerability shown by many implemented cryptographic
schemes \cite{Schneier98,Piramithu10,Naccache12}, the lack of a fundamental proof that establishes the difficulty of the decryption problem faced by adversaries, and the potential for transformative changes in computing motivate security solutions that are provably unbreakable.

After some initial theoretical studies by Wyner and Maurer, aspects of secrecy at the \emph{physical layer} have experienced a resurgence of interest only in the past decade or so. Therefore, the remainder of this paper is devoted to surveying and reviewing the various aspects of physical layer security in modern wireless networks.
The fundamental principle behind physical layer security is to exploit the inherent randomness of noise and communication channels to limit the amount of information that can be extracted at the `bit' level by an unauthorized receiver. More importantly, no limitations are assumed for the eavesdropper in terms of computational resources or network parameter knowledge, and the achieved security can be quantified precisely. With appropriately designed coding and transmit precoding schemes in addition to the exploitation of any available channel state information, physical layer security schemes enable secret communication over a wireless medium without the aid of an encryption key. However, if it is desirable to use a secret key for encryption, then information-theoretic security also describes techniques that allow for the evolution of such a key over wireless channels that are observable by the adversary. Thus, information-theoretic security is now commonly accepted as the strictest form of security. Additionally, since they can operate essentially independently of the higher layers, physical layer
techniques can be used to augment already existing security measures. Such a multilayered approach is expected to significantly enhance the security of modern data networks, whether wired or wireless.

Instead of proceeding in a strictly chronological order, we aim to provide a high-level overview of the historical development of the field along with the most pertinent references, juxtaposed with recent and ongoing research efforts. The foundations of single and multi-antenna wiretap channels are treated with some emphasis on the mathematical aspects, in order to facilitate the understanding of advanced multi-user networks. The term physical layer security will be used to encompass both signal processing and information-theoretic treatments of the topic.

The remainder of the article is organized as follows. In the next section, the fundamental mathematical precepts of secrecy are presented, along with a description of the most elementary secrecy problem: the wiretap channel. The state-of-the-art in the burgeoning area of multi-antenna wiretap channels is described in Section~\ref{sec:MIMOWiretap}. The extension to more than three terminals for broadcast, multiple-access, and interference channels is described in Section~\ref{sec:BC_MAC_IC}. The development of secrecy in relay channels and other cooperative scenarios is carried out in \ref{sec:Relay}. The important issue of secret-key generation and agreement in wireless networks is studied in Section~\ref{sec:Key}. Section~\ref{sec:CodeDesign} highlights the emerging areas of practical security based on error-correcting codes. The penultimate section covers cross-disciplinary approaches to secrecy based on game theory and stochastic geometry, miscellaneous systems such as sensor and cognitive radio networks, along with physical layer message authentication.
Finally, in Section~\ref{sec:concl} we summarize our discussion and provide a broad picture of future research directions.  Readers interested in going beyond the treatment of physical layer security offered in this paper are referred to the recent monographs \cite{LiangBook09}-\cite{JorsweickBook10}.


\section{Fundamentals}\label{sec:Fundamentals}
The simplest network where problems of secrecy and confidentiality arise is a three-terminal system comprising a transmitter, the intended (legitimate) receiver, and an unauthorized receiver, wherein the transmitter wishes to communicate a private message to the receiver. In the sequel, the unauthorized receiver is referred to interchangeably as an \emph{eavesdropper} or \emph{wiretapper}. The vast majority of physical layer security research reviewed in this survey contains the premise that the eavesdropper is passive, i.e., does not transmit in order to conceal its presence. The knowledge available to the transmitter regarding the eavesdropper's channel state information (CSI) plays a critical role in determining the corresponding optimal transmission scheme. Due to uncertainties regarding the location of eavesdroppers, this knowledge may range from a complete lack of CSI, to partial and statistical CSI, and all the way to complete CSI, as discussed in detail in the current and next section. Furthermore, knowledge of the statistical distributions of the eavesdropper spatial locations may also be beneficial, as discussed further in Sec.~\ref{sec:StochGeom}.

Encryption of messages via a secret key known only to the transmitter and intended receiver has been the traditional route to ensuring confidentiality. In the early 20th century, the design of cryptographic methods was based on the notion of computational security, without a solid mathematical basis for secrecy. A classical example was Vernam's one-time pad cipher \cite{Vernam26}, where the binary message or plaintext is XOR'ed with a random binary key of the same length.

\subsection{Performance Metrics}\label{sec:PerfMetrics}
Shannon postulated the information-theoretic foundations of modern cryptography in his ground-breaking treatise of 1949 \cite{Shannon49}. Shannon's model assumed that a non-reusable private key $K$ is used to encrypt the confidential message $M$ to generate the cryptogram $C$, which is then transmitted over a noiseless channel. The eavesdropper is assumed to have unbounded computational power, knowledge of the transmit coding scheme, and access to an identical copy of the signal at the intended receiver. The notion of perfect secrecy was introduced, which requires that the \emph{a posteriori} probability of the secret message computed by the eavesdropper based on her received signal be equal to the \emph{a priori} probability of the message. In other words, perfect secrecy implies
\begin{equation}\label{eq:Shannon_perfsecrecy}
I(M;C)=0,
\end{equation}
where $I(\cdot;\cdot)$ denotes mutual information.
A by-product of this analysis was that perfect secrecy \cite{Geffe65} can be guaranteed only if the secret key has at least as much entropy as the message to be encrypted (generally equivalent to the key and plaintext being of equal length \cite{HellmanMarch77}), i.e., $H(K) \geq H(M)$, which validated Vernam's one-time pad cipher system. In subsequent years, it became common practice to use the nomenclature Alice, Bob, and Eve to refer to the legitimate transmitter, intended receiver, and unauthorized eavesdropper, respectively.

\begin{figure}[htp]
\centering
\includegraphics[width=\linewidth]{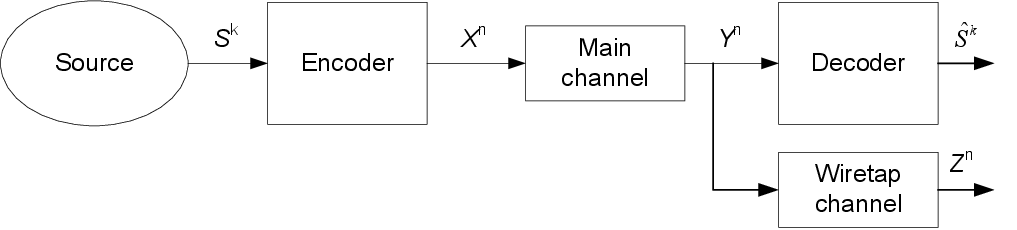}
\caption{The wiretap channel of Wyner \cite{Wyner75}, where the eavesdropper's discrete memoryless channel is degraded relative to the main channel.}
\label{fig_Wyner_Wiretap}
\end{figure}

Wyner ushered in a new era in information-theoretic security when he introduced the wiretap channel in \cite{Wyner75}, which considered the imperfections introduced by the channel. Here, the information signal $X$ is transmitted to the intended receiver Bob over the `main channel' which is modeled as a discrete memoryless channel. The receiver observes $Y$, which subsequently passes through an additional `wiretap channel' before being received by the eavesdropper as $Z$, as shown in Fig.~\ref{fig_Wyner_Wiretap}.

Under the assumption that the source-wiretapper link is a probabilistically degraded version of the main channel \cite{HellmanMarch77}, Wyner sought to maximize the transmission rate $R$ in the main channel while making negligible the amount of information leaked to the wiretapper. More specifically, the transmitter has a single message $W$, which is uniformly
distributed over $\{1,..., 2^{nR}\}$, where $R$ is the \emph{rate} of
communication and $n$ is the block length of communication. The goal of the transmitter is to deliver $W$
reliably to the legitimate receiver while keeping it secure from the eavesdropper. In the
classical work of \cite{Wyner75}, for every $\epsilon>0$  it is required that
\begin{equation}\label{eq:EquivocationRate}
R_e-\epsilon \leq \frac{1}{n}\,H(W\mid Z^n)
\end{equation}
for sufficiently large $n$, where
$R_e$ represents the uncertainty of message $W$ or the \emph{equivocation} at the eavesdropper \cite{RuohISIT2011}. The \emph{capacity-equivocation} region is then the set of rate-equivocation pairs $(R\,,\,R_e)$ that can be achieved by any coding scheme.

It is noted that $R-R_e = \frac{1}{n}\, I(W;Z^n)$ represents the information that is leaked to the eavesdropper. Thus, when the equivocation rate $R_e$ is arbitrarily close to the information rate $R$, message $W$ is asymptotically \emph{perfectly} secure
from the eavesdropper, i.e., \cite{RuohISIT2011}
\begin{align}\label{perfectCs}
\frac{1}{n}\, I(W;Z^n)\le \epsilon  \;.
\end{align}
Under the asymptotic perfect secrecy
constraint (\ref{perfectCs}), the maximum rate of communication $R$ is called
the \emph{secrecy capacity} of the wiretap channel. Also, it should be clear that the capacity of the direct link, without secrecy constraints, is the maximum rate $R$ in the capacity-equivocation region regardless of the value of $R_e$ and the secrecy constraint (\ref{eq:EquivocationRate}). This way, one induces maximal equivocation at the wiretapper, and Wyner was able to show that secure communication was possible \emph{without} the use of a secret key. Strictly speaking, Wyner's definition of ``perfect secrecy" as the scenario in
which the block-length-normalized mutual information at the eavesdropper vanishes in the limit of long block lengths was weaker than that proposed by Shannon [cf. \eqref{eq:Shannon_perfsecrecy}], which requires that the mutual information be zero regardless of the block length and is also known as strong secrecy \cite{Csizar96}.

More recently, the study of secrecy in fading channels has led to the use of outage probability performance metrics. Outage metrics for physical layer security are defined analogously to the conventional rate outage metrics, for e.g., the secrecy outage probability is the likelihood that the instantaneous secrecy rate $R_s$ is below a pre-defined threshold $\varepsilon$ for a particular fading distribution \cite{Barros06}:
\[{P_{out}} = \Pr \left\{ {{R_s} < \varepsilon } \right\}, \; \varepsilon > 0.\]
Furthermore, security approaches based on signal processing methods often make use of more traditional performance metrics by designing transmission schemes that restrict the bit error rate (BER) or signal-to-interference-plus-noise ratio (SINR) at eavesdroppers to pre-defined thresholds. Note that constraining the BER or SINR at eavesdroppers does not satisfy either weak or strong secrecy requirements, but can often simplify system design.

In 1993, Maurer \cite{Maurer93} presented a strategy that
allowed a positive rate even when the wiretapper observes a
``better" channel than the one used by the legitimate users. The essence of Maurer's scheme was the joint development of a secret key by the transmitter and receiver via communication over a public (insecure) and error-free feedback channel.
Thereafter, research in information-theoretic secrecy developed along two main branches: secret key-based secrecy as in the work by Shannon and Maurer, and keyless security as in the work by Wyner. In Section~\ref{subsec:SISOWiretap} to Section~\ref{sec:Relay} we trace the evolution of keyless security over the past four decades. We revisit the topic of key-based security for wireless channels in Section~\ref{sec:Key}.

\subsection{Single-Antenna Wiretap Channels Since Wyner}\label{subsec:SISOWiretap}
Early work in the field generally assumed non-fading channels, and knowledge of the (fixed) channel state was presumed at the transmitter. In \cite{Yakovlev81}, bounds on the equivocation rate for Wyner's wiretap channel model with finite code block lengths are derived. Carleial and Hellman \cite{Hellman77} considered a special case of Wyner's model where the main channel is noiseless and the wiretap channel is a binary symmetric channel, and analyzed the applicability of systematic linear codes for preserving the secrecy of an arbitrary portion of the transmitted message.
For the degraded wiretap channel \cite{LeungH78} with additive Gaussian noise, and $C_M$ and $C_W$ as the Shannon capacities of the main and wiretap channels, the essential result for the secrecy capacity $C_S$ was the following:
\begin{equation}\label{eq:degradedGaussianSecCap}
C_S=C_M-C_W.
\end{equation}
Ultimately, it was established that a non-zero secrecy capacity can only
be obtained if the eavesdropper's channel is of lower quality than
that of the intended recipient.

Csisz\`{a}r and K\"{o}rner considered a more general (non-degraded) version of Wyner's wiretap channel in \cite{Csiszar78}, where they obtained a single-letter characterization of the achievable \{private message rate,  equivocation rate, common message rate\}-triple for a two-receiver broadcast channel. For the special case of no common messages, the secrecy capacity was defined as
\begin{equation}\label{eq:Korner1978}
C_S  = \mathop {\max }\limits_{V \to X \to YZ} I\left( {V;Y} \right) - I\left( {V;Z} \right),
\end{equation}
which is achieved by maximizing over all joint probability distributions such that a Markov chain $V,X,YZ$ is formed, where $V$ is an auxiliary input variable.
In \cite{Mitrpant06} it was shown that the availability of non-causal side information at the encoder can enhance the achievable secrecy rate region of \eqref{eq:Korner1978}, based on dirty-paper coding arguments.

In \cite{Ozarow84}, Ozarow and Wyner studied the type-II wiretap channel, where the main communication
channel is noiseless but the wiretapper has access to an arbitrary subset $\mu$
of the $N$ coded bits, and optimal tradeoffs between code rate $k/N$ and $\mu$  that guaranteed secrecy were characterized.

The consideration of channel fading in wiretap channels has recently opened new avenues of research. Works in this area generally assume that at least the statistics of the eavesdropper's fading channel are known to the transmitter. Barros and Rodrigues \emph{et al.} \cite{Barros06}-\cite{Bloch06} analyzed the outage probability and outage secrecy capacity of slow
fading channels and showed that with fading, information-theoretic security is achievable even
when the eavesdropper has a better average SNR than the legitimate receiver.

Li \emph{et al}. \cite{Trappe07} examined the achievable secrecy rate for an AWGN main channel and a Rayleigh fading eavesdropper's channel with additive Gaussian noise, assuming that the eavesdropper channel realizations
are unknown to legitimate transmitter Alice and receiver Bob. The main result of this paper was
that with Gaussian random codes, artificial noise injection and
power bursting, a positive secrecy rate is achievable even when
the main channel is arbitrarily worse than the eavesdropper's average channel. A more exotic scenario was studied in \cite{Ozel12} where the source has a stochastic power supply based on energy harvesting. Here, the i.i.d. energy arrivals are equated to channel states that are known causally to the source, and the optimal input distribution that attains the boundary of the
capacity-equivocation region of the Gaussian wiretap channel was derived. Here, the capacity corresponds to the reliability of the main channel, while the equivocation refers to the normalized conditional entropy at the eavesdropper as described in Sec.~\ref{sec:Fundamentals}.

Relatively fewer studies consider the case of a complete absence of eavesdropper CSI at the transmitter in fading wiretap channels. In \cite{ElGamal08}, the authors considered a block-fading scalar wiretap channel where the number of channel uses within each coherence interval is large enough to invoke random coding arguments. This assumption is critical for their achievable coding scheme which attempts to ``hide" the secure message across different fading states.
A recent approach towards understanding the information-theoretic limits of wiretap channels with no eavesdropper CSI has been taken by studying the compound wiretap channel \cite{Kramer09}. The compound wiretap channel captures the situation in which there is no or incomplete CSI at the transmitter by assuming the eavesdropper's channel is always drawn from a finite, known set of states, and guarantees secure communication under any state that may occur.

\section{Multi-antenna Channels}\label{sec:MIMOWiretap}
\begin{figure}[htp]
\centering
\includegraphics[width=\linewidth]{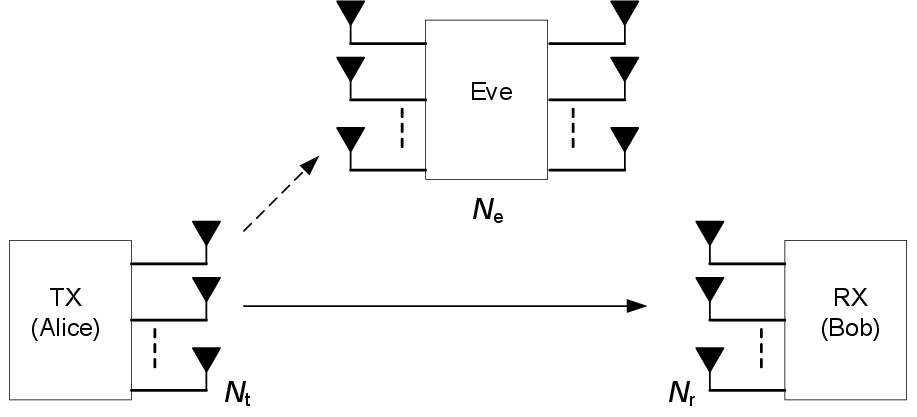}
\caption{General MIMO wiretap channel.}
\label{fig_MIMOwiretap}
\end{figure}

The explosion of interest in multiple-input multiple-output (MIMO) systems soon led to the realization that exploiting the available spatial dimensions could also enhance the secrecy capabilities of wireless channels. In a fading MIMO channel where the transmitter, receiver, and eavesdropper are equipped with $N_T,N_R,N_E$ antennas respectively as in Fig.~\ref{fig_MIMOwiretap}, a general representation for the signals received by the legitimate receiver and passive eavesdropper are
\begin{equation}\label{eq:yHe}
\begin{split}
\mathbf{y}_b  &=  \mathbf{H}_{b} \mathbf{x}_a +  \mathbf{n}_b \\
\mathbf{y}_e  &=  \mathbf{H}_{e} \mathbf{x}_a + \mathbf{n}_e,
\end{split}
\end{equation}
where $\mathbf{x}_a \in \mathbb{C}^{N_T\times 1}$ is the transmit signal with covariance matrix $E\left\{ {{\mathbf{x}}_a {\mathbf{x}}_a^H } \right\} = {\mathbf{Q}}_x$, average power constraint $\operatorname{Tr} \left( {{\mathbf{Q}}_x } \right) \leq P$, $\mathbf{H}_{b} \in \mathbb{C}^{N_R\times N_T}, \mathbf{H}_{e} \in \mathbb{C}^{N_E\times N_T}$ are the MIMO complex Gaussian channel matrices, and $\mathbf{n}_b, \mathbf{n}_e$ are zero-mean complex white Gaussian additive noise vectors.

The work by Hero \cite{Hero03} was arguably the first to consider secret communication in a MIMO setting, and sparked a concerted effort to apply and extend the single-antenna wiretap theory to this new problem. Hero examined the utility of space-time block coding for covert communications in \cite{Hero03}, and designed CSI-informed transmission strategies to achieve either a low probability of intercept (defined in terms of eavesdropper mutual information), or a low probability of detection for various assumptions about the CSI available to the eavesdropper. One of the  main results was that if the eavesdropper is completely unaware of its receive CSI, then an equivocation-maximizing strategy is to employ a space-time constellation with a constant spatial inner product.

Parada and Blahut analyzed a degraded single-input multiple-output (SIMO; $N_T=1, N_R,N_E>1$) wiretap channel in \cite{Blahut05}, and obtained a single-letter characterization of its secrecy capacity by transforming the problem to a scalar Gaussian wiretap channel and then re-applying (\ref{eq:degradedGaussianSecCap}). The authors also proposed a secrecy rate outage metric for the SIMO wiretap channel with slow fading, and observed a secrecy diversity gain of order proportional to the number of receiver antennas. The corresponding multiple-input single-output (MISO) case was studied in \cite{Yates07,Shafiee07}, where it was noted that the MIMO wiretap channel is not degraded in general. Since this renders a direct computation of (\ref{eq:Korner1978}) difficult, they therefore restricted attention to Gaussian input signals. For the special case of $N_T=2, N_R=2, N_E=1$ analyzed by Shafiee and coworkers in \cite{Shafiee09}, a beamforming transmission strategy was shown to be optimal.

The next steps toward understanding the full-fledged MIMO wiretap channel were taken in \cite{Khisti07}-\cite{GoelN08}, which considered the case of multiple antennas at all nodes and termed it the MIMOME (multiple-input multiple-output multiple-eavesdropper) channel. Khisti \emph{et al}. \cite{Khisti07} developed a genie-aided upper bound for the MIMO secrecy capacity for which Gaussian inputs are optimal.
 When the eavesdropper's instantaneous channel state is known at the transmitter, it was shown that an asymptotically optimal (high SNR) scheme is to apply a transmit precoder based upon the generalized singular value decomposition (GSVD) of the pencil $(\mathbf{H}_b,\mathbf{H}_e)$, which decomposes the system into parallel channels and leads to a closed-form secrecy rate expression. For the so-called MISOME special case where $N_R=1, N_T,N_E>1$, the optimal transmit beamformer is obtained as the generalized eigenvector ${\bm{\psi }}_m$ corresponding to the largest generalized eigenvalue $\lambda _m$ of
 \[
{\mathbf{h}}_b^H {\mathbf{h}}_b {\bm{\psi }}_m  = \lambda _m {\mathbf{H}}_e^H {\mathbf{H}}_e {\bm{\psi }}_m.
\]

 If only the statistics of $\mathbf{H}_e$ are known to the transmitter, then the authors proposed an \emph{artificial noise} (AN) injection strategy as first suggested by Goel and Negi \cite{Negi05,GoelN08}. The artificial noise is transmitted in conjunction with the information signal, and is designed to be orthogonal to the intended receiver, such that only the eavesdropper suffers a degradation in channel quality \cite{Khisti10,Wornell09}. The transmit signal can be represented in general as
 \begin{equation}
{{\mathbf{x}}_a} = {{\mathbf{T}}_a}{{\mathbf{z}}_a} + {{\mathbf{T}}_n}{{\mathbf{z}}_n}
 \end{equation}
where precoding matrices ${{\mathbf{T}}_a} \in {\mathbb{C}^{{N_T} \times {N_T} - d}}$ and ${{\mathbf{T}}_n}\in {\mathbb{C}^{{N_T} \times d}}$ correspond to data and AN signal vectors $\mathbf{z}_a\in {\mathbb{C}^{ {N_T} - d\times 1}}$, $\mathbf{z}_n\in {\mathbb{C}^{ d\times 1}}$, respectively. When $N_T>N_R$, ${{\mathbf{T}}_n}$ can be formed from the nullspace of $\mathbf{H}_{b}$, otherwise ${{\mathbf{T}}_n}$ and ${{\mathbf{T}}_a}$ can be chosen to guarantee received signals in orthogonal spaces by forming them from the right singular vectors of $\mathbf{H}_{b}$ \cite{Mukherjee09}. If the eavesdropper's CSIT is partially known, additional gains may be achieved by optimizing the AN transmit covariance \cite{WingKMa13} or relaxing the orthogonality constraint \cite{Su13}. As will be seen in the rest of the survey, the use of artificial noise is a recurring theme for secrecy in many different multiuser networks.

\begin{figure}[htp]
\centering
\includegraphics[width=\linewidth]{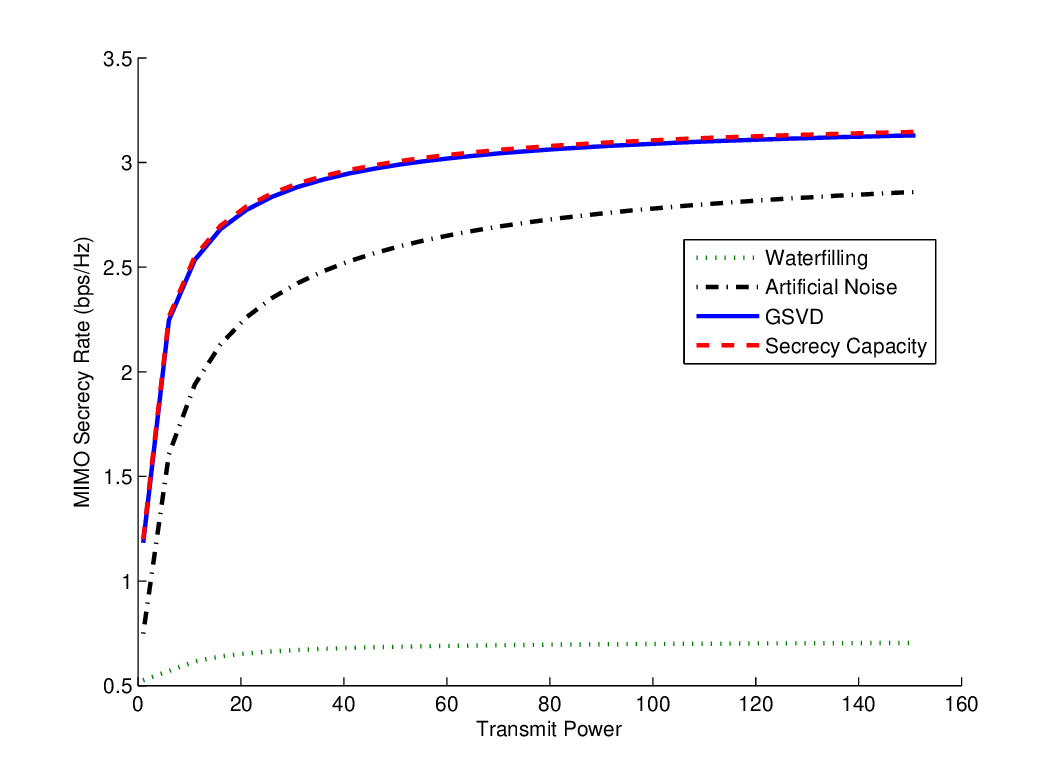}
\caption{The MIMO secrecy rates of GSVD-beamforming \cite{Ali10,Khisti10}, artificial noise \cite{GoelN08}, and waterfilling over the main channel, $N_T=N_E=3,N_R=2$. Transmit power is in dB, assuming 0dB noise power.}
\label{fig_MIMOSecrecy}
\end{figure}
An example of the secrecy rate performance of various transmission strategies for the MIMO wiretap channel is shown in Fig.~\ref{fig_MIMOSecrecy}. The GSVD scheme requires instantaneous knowledge of eavesdropper channel $\mathbf{H}_e$, the artificial noise scheme requires the statistics of $\mathbf{H}_e$, and the relatively poor performance of waterfilling on the main channel is also shown when no information is available regarding $\mathbf{H}_e$.

The MIMO wiretap channel was studied independently by Oggier and Hassibi \cite{OggierH08}, who computed a similar upper bound on the MIMO secrecy capacity,
and showed after a matrix optimization analysis that
\begin{equation}\label{eq:MIMOsecCap}
C_S=\mathop {\max }\limits_{{\mathbf{Q}}_x  \succeq 0} \log \det \left( {{\mathbf{I}} + {\mathbf{H}}_b {\mathbf{Q}}_x {\mathbf{H}}_b^H } \right) - \log \det \left( {{\mathbf{I}} + {\mathbf{H}}_e {\mathbf{Q}}_x {\mathbf{H}}_e^H } \right).
\end{equation}
In \cite{LiuShitz09}, Liu and Shamai reexamine the MIMO wiretap channel with a more general matrix input power-covariance constraint ${\mathbf{Q}}_x  \preceq {\mathbf{S}}$, and showed that the conjecture of a Gaussian input $U = X$ without prefix coding is indeed an optimal secrecy capacity-achieving choice.
Zhang \emph{et al}. attempt to bypass the non-convex optimization of the optimal input covariance matrix by drawing connections to a sequence of convex cognitive radio transmission problems, and obtained upper and lower bounds on the MIMO secrecy capacity \cite{Cui10}. Li and Petropulu \cite{Petropulu10} computed the optimal input covariance matrix for a MISO wiretap channel, and presented a set of equations characterizing the general MIMO solution.

Bustin and coauthors \cite{Bustin09} exploited the fundamental relationship between mean-squared error and mutual information to provide a closed-form expression for the optimal input covariance $\mathbf{Q}_x$ that achieves the MIMO wiretap channel secrecy capacity, again under an input power-covariance constraint.
More precisely, it was shown in \cite{Bustin09} that, under the matrix power constraint $\mathbf{Q}_x\preceq \mathbf{S}$, the solution of (6) is given by
\begin{equation}\label{eq:MIMOsecCapunderS}
C_{sec}(\mathbf{S})=\sum_{i=1}^{\lambda}\log\alpha_i
\end{equation}
where $\alpha_i$, $i=1,\ldots,\lambda$, are the generalized eigenvalues of the pencil
\begin{equation}\label{eq:r7}
(\mathbf{S}^{\frac{1}{2}}{\mathbf{H}}_b^H{\mathbf{H}}_b\mathbf{S}^{\frac{1}{2}}+\textbf{I},\quad \mathbf{S}^{\frac{1}{2}}{\mathbf{H}}_e^H{\mathbf{H}}_e\mathbf{S}^{\frac{1}{2}}+\textbf{I})
\end{equation}
that are greater than 1.
Note that, since both elements of the pencil (\ref{eq:r7}) are strictly positive definite, all the generalized eigenvalues of the pencil (\ref{eq:r7}) have real positive values \cite{LiuLiu10,Horn}.
In (\ref{eq:MIMOsecCapunderS}), a total of $\lambda$ of them are assumed to be greater than 1. Clearly, if there are no such eigenvalues, then the information signal received at the intended receiver is a degraded version of that of the eavesdropper, and in this case the secrecy capacity is zero.

It should be noted that, under the average power constraint $\operatorname{Tr} \left( {{\mathbf{Q}}_x } \right) \leq P$, there is no computable secrecy capacity expression for the general MIMO case. In fact, for the average power constraint, the secrecy capacity would in principle be found through an exhaustive search over the set $\{\mathbf{S}: \mathbf{S}\succeq 0, \text{Tr}(\mathbf{S})\leq P\}$. More precisely, we have
\cite{LiuLiu10}, \cite[Lemma 1]{BCWeingarten06}
\begin{eqnarray}\label{eq:12}
C_{sec}(P)=\max_{\mathbf{S}\succeq 0, \text{Tr}(\mathbf{S})\leq P}C_{sec}(\mathbf{S})
\end{eqnarray}
where, for any given semidefinite $\mathbf{S}$, $C_{sec}(\mathbf{S})$ can be computed as given by (\ref{eq:MIMOsecCapunderS}). A closed-form solution is possible in certain special cases, for example when $\textbf{S}$ is known
to be full rank \cite{FakoorianTSP13,Charalambous12}, or in the high-SNR regime based on the GSVD \cite{Khisti10} as described previously.

Subsequently, numerous research contributions emerged that considered a number of practical issues regarding the MISO/MIMO wiretap channel \cite{HongMag13}, of which we enumerate a few below:
\begin{itemize}
\item Optimal power allocation and beamforming methods for the artificial noise strategy were presented in \cite{Zhou09}, for the MISO scenario in \cite{Ghogho12,Ghogho13,Jorswieck12}, and for the GSVD-based precoding scheme in \cite{Ali10}.
\item If even statistical information regarding the eavesdropper's channel is unavailable, then Swindlehurst \emph{et al}. \cite{Swindlehurst09,Mukherjee09} suggested an approach where just enough power is allocated to meet a target performance criterion (SNR or rate) at the receiver, and any remaining power is used for broadcasting artificial noise, since the secrecy rate cannot be computed at the transmitter. A compound wiretap channel approach and a resultant universal coding scheme that guarantees a positive secrecy rate was presented in \cite{He10}.
\item The effects of imperfect and quantized CSIT of the main (Alice-to-Bob) channel upon the secrecy rate were examined in \cite{Swindle10} and \cite{ISIT09}, respectively, while bounds on secrecy capacity with imperfect CSIT and limited ARQ feedback were given in \cite{AlouiniAsilomar11,AlouiniISIT12}. MIMOME secrecy rate maximization with imperfect CSIT of all channels was solved using an iterative algorithm in \cite{Cumanan14} via a Taylor series expansion to convexify the secrecy rate. Discriminatory training methods that include artificial noise for acquisition of main channel CSI while degrading the eavesdropper's estimate of $\mathbf{H}_e$ were analyzed in \cite{Hong13}.
\item Precoding and receive filter designs to minimize the mean-square error (MSE) at Bob while constraining the MSE at Eve to be above some threshold were given in \cite{Reboredo13}. Non-linear precoding based on lattices or vector-perturbation ideas with eavesdropper error probability as the metric was examined in \cite{Viterbo13}.
\item MIMO secrecy capacity has also been studied for OFDM-based frequency-selective channels \cite{Debbah08,Renna12}, Rician fading channels \cite{PetropuluRician11}, and ergodic \cite{Bhargava09} channel fading processes. The secrecy outage probability of maximum ratio combining was presented in \cite{Wang11} and of transmit antenna selection in \cite{Elkashlan13}-\cite{Latva-aho13}.
\item Detection-theoretic methods for discerning the presence of a completely passive eavesdropper based on its local oscillator leakage power were analyzed in \cite{MukherjeeICASSP12}.
\item An evolved full-duplex eavesdropper that can divide its antenna array into sub-arrays for simultaneous eavesdropping and jamming was considered in \cite{MukherjeeASILOMAR11}.
\end{itemize}

A summary of transmission strategies in the MIMO wiretap channel for various assumptions regarding eavesdropper channel state information at the transmitter (ECSIT) is presented in Table~\ref{table_MIMOstrats}.
\begin{table}[hbp]
\renewcommand{\arraystretch}{1.3}
\caption{Comparison of MIMO wiretap transmission strategies for various ECSIT assumptions}
\label{table_MIMOstrats}
\centering
\begin{tabular}{|l|l|l|}
\hline
\bfseries Parameters & \bfseries Strategy &\bfseries Criterion\\
\hline
MIMOME, no ECSIT \cite{Mukherjee09}       & Artif. noise    &Meet rate target\\
MIMOME, statistical ECSIT \cite{GoelN08}  & Artif. noise    &Ergodic secrecy rate\\
MISOME, complete ECSIT  \cite{Khisti10}   & GEVD            &Secrecy rate\\
MIMOME, complete ECSIT  \cite{Wornell09}  & GSVD            &Secrecy rate\\
\hline
\end{tabular}
\end{table}

\section{Broadcast, Multiple-Access, and Interference Channels}\label{sec:BC_MAC_IC}
\subsection{Broadcast and Multiple-Access Channels}
The concept of information-theoretic security is easily extended to larger multi-user networks with more than two receivers and/or transmitters.
We begin with one-to-many broadcast channels (BCs), which can be divided into two major categories from a security perspective:
\begin{enumerate}
\item BC with confidential messages: each downlink message must be kept confidential from all other unintended receivers, i.e., each receiver is seen as an eavesdropper for messages not intended for it.
\item Wiretap BC: messages do not need to be mutually confidential among the downlink receivers, but must be protected from external eavesdroppers.
\end{enumerate}
The former case is more challenging than the latter, for which the existing transmission techniques of Sec.~\ref{sec:MIMOWiretap} can mostly be reused. Therefore, unless stated otherwise the following discussion will assume the first category.

The original wiretap channel as proposed by Wyner \cite{Wyner75}, is a form of broadcast channel (BC) where the source sends confidential messages to the destination, and attempts to keep the messages as secret as possible from the other receiver(s)/ eavesdropper(s). Csisz\`{a}r and K\"{o}rner extended
this work to the case where the source sends common information to both the destination and the eavesdropper, and confidential messages are sent
only to the destination \cite{Csiszar78}. The secrecy capacity region of this scenario, for the case of a BC with parallel independent subchannels, was considered in \cite{Liang08} and the optimal source power allocation that achieves the boundary of the secrecy capacity region was derived. The secrecy capacity region of the MIMO Gaussian broadcast channel with common message to both the destination and the eavesdropper, and confidential message sent
only to the destination, was characterized in \cite{BCLiang09} using a channel enhancement approach \cite{BCWeingarten06} and under the matrix input power-covariance constraint ${\mathbf{Q}}_x  \preceq {\mathbf{S}}$. The notion of an enhanced broadcast channel was first introduced in \cite{BC Weingarten06} and was used jointly with the entropy power inequality to characterize the capacity region of the conventional Gaussian MIMO broadcast channel (without secrecy constraint).  Most of the current work in the literature on secrecy for the MIMO broadcast channel uses this idea. Moreover, instead of the average total power constraint $\operatorname{Tr} \left( {{\mathbf{Q}}_x } \right) \leq P$, they considered the matrix input power-covariance constraint ${\mathbf{Q}}_x  \preceq {\mathbf{S}}$.

The discrete memoryless broadcast channel with two confidential messages sent to two receivers, where each receiver acts as an eavesdropper for the other, was studied in \cite{Yates08}, where inner and outer bounds for the secrecy capacity region were established. This problem was studied in \cite{Liu09} for the MISO Gaussian case and in \cite{LiuLiu10} for the general MIMO Gaussian case. Rather surprisingly, it was shown in \cite{LiuLiu10} that, under the matrix input power-covariance constraint, both confidential messages can be simultaneously communicated at their respected maximum secrecy rates, where the achievability was obtained using dirty-paper coding. To prove this result, Liu \emph{et al}. revisited the MIMO Gaussian wiretap channel and showed that a coding scheme that uses artificial noise and random binning achieves the secrecy capacity of the MIMO Gaussian wiretap channel as well \cite{LiuLiu10}.

Consider the broadcast channel represented by \eqref{eq:degradedGaussianSecCap} and \eqref{eq:Korner1978}, with the addition of independent confidential messages $W_1$ (intended for receiver 1 but needed to be kept secret from receiver 2) and $W_2$ (intended for receiver 2 but needed to be kept secret from receiver 1). From \cite [Corollary 2]{LiuLiu10}, under the matrix constraint $\mathbf{S}$, the secrecy capacity region is given by the set of nonnegative rate pairs $(R_1,R_2)$ such that
\begin{equation}\label{eq:MIMOBCregS1}
R_1\leq\sum_{i=1}^{\lambda}\log\alpha_i ;\quad R_2\leq\sum_{j=1}^{N_T-\lambda}\log\frac{1}{\beta_j}
\end{equation}
where $\alpha_i$, $i=1,\ldots,\lambda$, are the generalized eigenvalues of the pencil (\ref{eq:r7}) that are bigger than 1, and $\beta_j$ $j=1$, \ldots,$(N_T-\lambda)$ are those that are less than or equal to 1.

The secrecy capacity region of MIMO Gaussian broadcast channels with confidential and common messages, where the transmitter has two independent confidential messages and a common message, was characterized in \cite{BCLiuISIT10}. The achievability was obtained using secret dirty-paper coding, while the converse was proved by using the notion of channel splitting \cite{BCLiuISIT10}.
Secure broadcasting with more than two receivers was considered in \cite{Wornell08}--\cite{BCLiuLiu10} (and references therein). These papers assume one transmitter intends to communicate with several legitimate users in the presence of an external eavesdropper. The secrecy capacity region for the case of two legitimate receivers was characterized by Khandani \emph{et al.} \cite{BCKhandani09} using enhanced channels, and for an arbitrary number of legitimate receivers by Ekrem \emph{et al.} \cite{BCUlukus09}. Ekrem \emph{et al.} use the relationships between minimum-mean-square-error and mutual information, and equivalently, the relationships between Fisher information and differential entropy to provide the converse proof.
In \cite{BCLiuLiu10}, Liu \emph{et al.} considered the secrecy capacity regions of the degraded Gaussian MIMO BC with layered confidential messages, where each message must be kept secret from different subsets of receivers. They presented a vector generalization of Costa's Entropy Power Inequality to provide their converse proof. Chia and El Gamal provided inner and outer bounds on the secrecy capacity region of
the three-receiver BC with one common and one confidential
message in \cite{ChiaGamal12}, and the extension to additional layered message sets was studied in \cite{Aref13}. The role of artificial noise for jamming eavesdroppers in wiretap broadcast channels was investigated in \cite{MukherjeeAllerton09,ICASSP10,Ng12}.

For the average transmit power constraint $\operatorname{Tr} \left( {{\mathbf{Q}}_x } \right) \leq P$, a computable
secrecy capacity expression is currently unavailable for the general MIMO broadcast channel case. However, optimal
solutions based on linear precoding have been found. For example, in \cite{Fakoorian13}, a linear precoding
scheme was proposed for a general MIMO BC under the matrix covariance constraint. Conditions were derived under which the proposed linear precoding approach is optimal and achieves the same secrecy rate region as S-DPC. This result was then used to derive a
closed-form sub-optimal algorithm based on linear precoding for an average power constraint.
In \cite{Fakoorian11}, GSVD-based beamforming was used for the MIMO Gaussian BC to simultaneously
diagonalize the channels. Linear precoding based on regularized channel inversion was studied in \cite{Collings12,Collings13,CollingsCOMML13} for a multi-antenna downlink where each message must be kept confidential from unintended receivers, and additional external eavesdroppers were assumed present in \cite{Geraci13}.  User selection in downlink channels with external eavesdroppers was studied in
\cite{MukherjeeAsilomar09}--\cite{PeiCOMML14}.

Other recent work on secure multi-user communications investigate the multiple-access
channel (MAC) with confidential messages \cite{ MACMaric06}, \cite{LiangIT08}, the
MAC wiretap channel (MAC-WT) \cite{Tekin08a}, \cite{Tekin08b}, and the cognitive MAC with confidential messages \cite{MACLiu09}. In \cite{ MACMaric06} and \cite{LiangIT08}, two transmitters communicating with a common receiver try to keep their messages secret from each other. For this scenario, the achievable secrecy rate region, and the capacity region for some special cases, are considered.

In \cite{Tekin08a}, the Gaussian multiple access
wire-tap channel (GMAC-WT) was considered, where multiple users are transmitting
to a base station in the presence an eavesdropper that receives
a noisy version of what is received at the base station (degraded wiretapper). In \cite{Tekin08a}, achievable rate regions were found
for different secrecy constraints, and it was shown that the secrecy
sum capacity can be achieved using Gaussian inputs and
stochastic encoders.
 In \cite{Tekin08b,Tekin08c}, a general, not necessarily degraded, Gaussian
MAC-WT was considered, and the optimal transmit power
allocation that achieves the maximum secrecy sum-rate was obtained.
It was shown in \cite{Tekin08b} that, a user that is prevented from transmitting based on the obtained power allocation can help increase the secrecy rate for
other users by transmitting artificial noise to the eavesdropper.

In \cite{MACLiu09}, Liu \emph{et al}. considered the fading cognitive multiple-access channel with confidential messages (CMAC-CM), where two users attempt to transmit common information to a destination while user 1 also has confidential information intended for the destination and tries to keep its confidential messages as secret as possible from user 2. The secrecy capacity region of the parallel CMAC-CM was established and the closed-form power allocation that achieves every boundary point of the secrecy capacity region was derived \cite{MACLiu09}. It should be noted that all the above work on the MAC with confidential messages assumes single antenna nodes, with little existing work on multiple-antenna scenarios.

\subsection{Interference Channel}\label{sec:IC}
The interference channel (IFC) refers to the case where multiple
communication links are simultaneously active in the same time and
frequency slot, and hence potentially interfere with each other. The IC is generally considered to be the antithesis of a cooperative network, since each transmitter is interested only in selfishly maximizing its own rate, and its message acts as interference to all other links. In conventional IFCs it is generally assumed that each receiver treats the interference from unintended transmitters as noise, but under secrecy constraints this assumption can no longer be made. A special application of the IFC with secrecy constraints is addressed
in \cite{Shitz09}, where the message from only one of the transmitters was
considered confidential.  The more general case, where each receiver acts as an eavesdropper for the other transmitter, was studied in \cite{Yates08} where, in the absence of a common message, the authors imposed a perfect secrecy constraint and obtained inner and outer bounds for the
perfect secrecy capacity region. In \cite{Chandramouli10}, the authors analyzed the optimal location of an external eavesdropper so as to drive the secrecy rate of all links to zero, where location is defined logically in terms of channel gains.

Since in most multi-user scenarios it is difficult to obtain
the exact secrecy capacity region, there has been recent interest in studying the asymptotic performance of
these systems in the high SNR regime. For such networks, a useful metric that captures the scaling behavior of the sum secrecy rate $R_\Sigma$ as the transmit SNR, $\rho$, goes to infinity is the number of secure degrees of freedom (SDoF), which can be defined as
\[
\eta  \triangleq \mathop {\lim }\limits_{\rho  \to \infty } \frac{{R_\Sigma  \left( \rho  \right)}}
{{\log \left( \rho  \right)}}.
\]
The SDoF of various multiuser networks described in Sections~\ref{sec:BC_MAC_IC}-\ref{sec:Relay} are summarized in Table~\ref{table_SecureDoF}, and generally rely upon the principle of interference alignment (IA) for achievability \cite{Jafar09}. For example, the number of secure DoF for $K$-user Gaussian IFCs ($K\geq 3$) has been addressed in \cite{Koylouglu08}, \cite{Koylouglu09}, \cite{IFCYener09}, and it was shown that under very strong
interference, positive secure DoFs are achievable via IA and channel extension. The $(K \times L)$ $X$ network comprises $K$ transmitters that each wish to communicate with $L$ receivers, and each of the receivers wishes to receive messages from all $K$ transmitters, and the SDoF is achieved via random binning and IA \cite{Jafar08}.
\begin{table}[hbtp]
\renewcommand{\arraystretch}{1.3}
\caption{Secure degrees of freedom in multiuser networks.}
\label{table_SecureDoF}
\centering
\begin{tabular}{|l|l|}
\hline
\bfseries Network & \bfseries Secure DoF\\
\hline
$K$-user SISO IFC, confidential messages \cite{Koylouglu08}       & $\eta=\frac{K-2}{2K-2}$\\
$K$-user SISO IFC, external Eve  \cite{Koylouglu08}               & $\eta=\frac{K-2}{2K}$\\
$K$-user SISO MAC, external Eve \cite{UlukusISIT13}               & $\eta=\frac{K(K-1)}{K(K-1)+1}$\\
$K$-helper SISO wiretap, external Eve \cite{UlukusCISS13}         & $\eta=\frac{K}{K+1}$\\
$(K \times L)$ $X$ network, confidential messages \cite{Jafar08}  & $\eta=\frac{L(K-1)}{K+L-1}$\\
\hline
\end{tabular}
\end{table}

It should be noted that all of the above
references \cite{Shitz09}-\cite{ IFCYener09} assume single antenna nodes. The more limited set of work that considers the impact of multi-antenna nodes on secrecy in the interference channel include \cite{IFCJorswieck09}-\cite{IFCAlij10}. In \cite{IFCJorswieck09}, Jorswieck \emph{et al}. studied the achievable secrecy rates of a two-user MISO interference channel, where each receiver has a single antenna. They modeled a non-cooperative game in the MISO interference channel and obtained the Nash equilibrium point using an iterative algorithm. A more unusual formulation was adapted in \cite{FakoorianTSP13MISO}, where a closed-form solution for the NE point was obtained where each multi-antenna transmitter desires to maximize the difference
between its secrecy rate and the secrecy rate of the other link.

\begin{figure}[htbp]
\centering
\begin{subfigure}[b]{0.7\textwidth}
\includegraphics[width=3.1in]{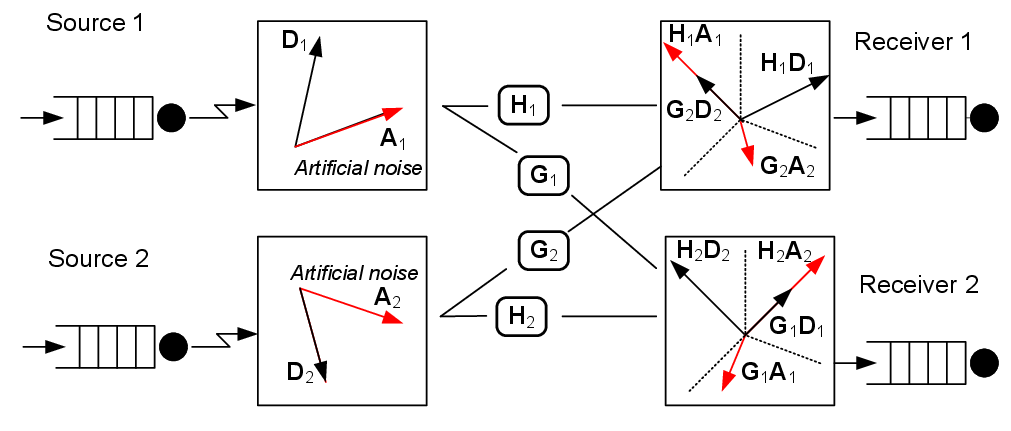}
\end{subfigure}
~
\begin{subfigure}[b]{0.35\textwidth}
\includegraphics[width=2.2in]{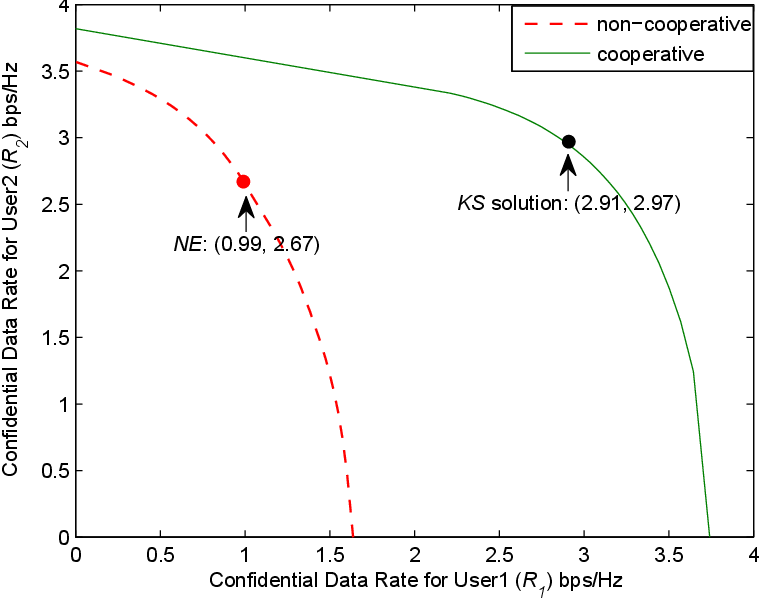}
\end{subfigure}
\caption{The cooperative ANA principle in the 2-user MIMO interference channel, and the corresponding secrecy rate region with and without cooperation.}
\label{fig:ANA}
\end{figure}

In \cite{IFCAlic10} and \cite{IFCAlij10}, Swindlehurst \emph{et al}. investigated the two-user MIMO Gaussian interference channel with confidential messages, where each node has arbitrary number of antennas. Several cooperative and non-cooperative transmission schemes were described, and their achievable secrecy rate regions were derived. A game-theoretic formulation of the problem was adopted to allow
the transmitters to find an operating point that balances network performance and fairness (the so called Kalai-Smorodinsky (K-S) bargaining solution \cite{IFCAlij10}). If the transmitters cooperate by exchanging information about the channels and signal subspaces associated with their link, then a combination of GSVD beamforming and altruistic \emph{artificial noise alignment} by each transmitter to mask the information
signal from the other transmitter at its \emph{own} receiver can be used, as seen in Fig.~\ref{fig:ANA}. As depicted in the figure, each transmitter intentionally undermines the ability of its receiver to decode the interfering signal; for example, noise $\mathbf{H}_1 \mathbf{A}_1$ and interference $\mathbf{G}_2 \mathbf{D}_2$ are aligned to lie in the same subspace at receiver 1. Here, the artificial noise can potentially also degrade the confidential message of the transmitter itself, so the transmit signal and power allocated to noise must be carefully designed. It was shown in \cite{IFCAlij10} that, while ordinary jamming is near optimal for the
standard wiretap channel \cite{Wornell09}, its performance is far from optimal for the interference channel.
Fig.~\ref{fig:ANA} shows the achievable secrecy rate regions of the proposed schemes in \cite{IFCAlij10} with 2 antennas at source 1, 3 antennas at all other nodes, and a transmit SNR of 20 dB, along
with the Nash equilibrium (NE) from the non-cooperative GSVD approach, and the clearly superior K-S rate point for the
cooperative GSVD and artificial noise alignment method.

\section{Relays and Cooperative Methods}\label{sec:Relay}
\begin{figure}[htbp]
\centering
\begin{subfigure}[b]{0.5\textwidth}
\includegraphics[width=3.2in]{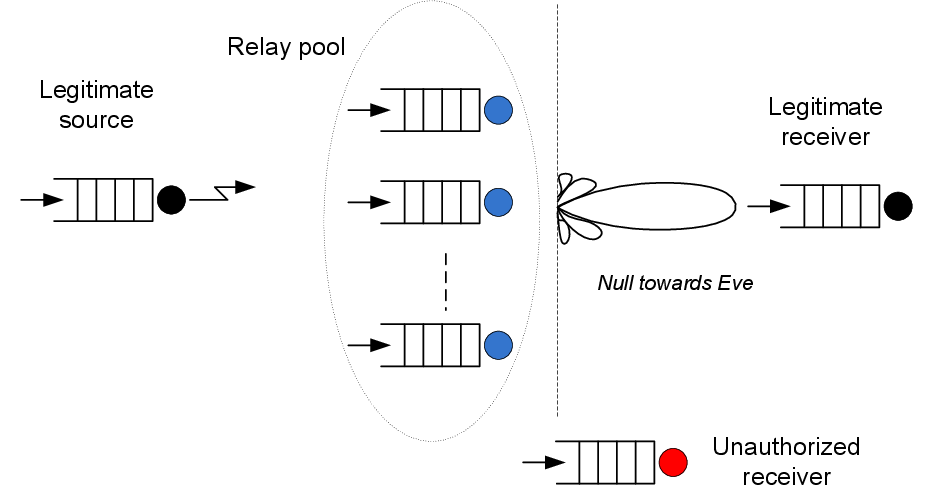}
\caption{Secure collaborative beamforming with nulls directed towards Eve.}
\label{fig_CollabBF}
\end{subfigure}
~\quad
\begin{subfigure}[b]{0.5\textwidth}
\includegraphics[width=3.2in]{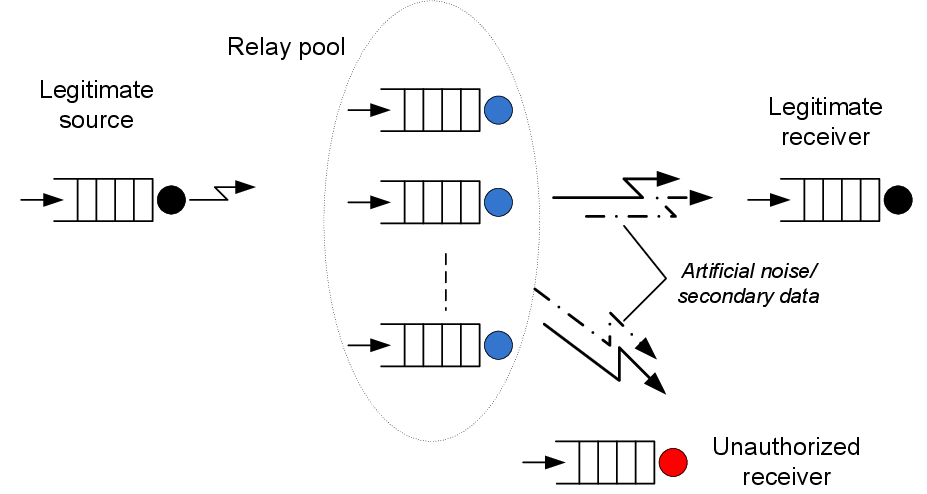}
\caption{Cooperative jamming of eavesdropper with artificial noise.}
\label{fig_CollabBF}
\end{subfigure}
~
\begin{subfigure}[b]{0.5\textwidth}
\includegraphics[width=3.2in]{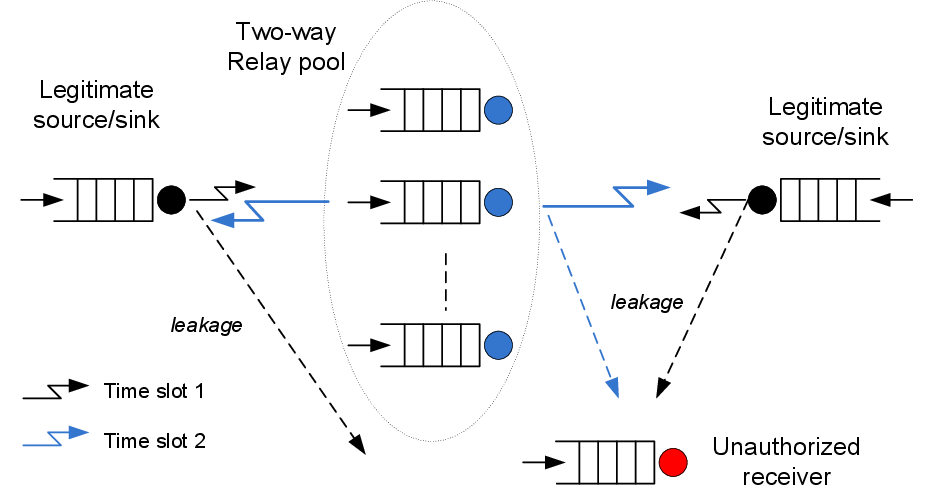}
\caption{Two-way relay-aided secret key exchange based on analog network coding.}
\label{fig_CollabBF}
\end{subfigure}
\caption{Relay-aided cooperative approaches for physical layer security with an external eavesdropper.}
\label{fig_mix}
\end{figure}

The issue of physical layer security in relay and cooperative networks has drawn much attention recently, as a natural extension to the secure transmission problem in non-cooperative networks. The secrecy capacity and achievable secrecy rate bounds have been investigated for various types of relay-eavesdropper channels, and many cooperative strategies stemming from conventional relay systems have been adopted with modifications based on techniques discussed in Sec.~\ref{sec:MIMOWiretap}, as shown in Fig.~\ref{fig_mix}. Security issues in relay networks can be divided into two broad categories:
\begin{itemize}
\item Relays are untrusted nodes from whom the transmitted messages must be kept confidential even while using them to relay those messages,
\item Relays are trusted nodes from whom the transmitted messages need not be kept secret.
\end{itemize}

\begin{figure}[htbp]
\centering
\includegraphics[width=0.4\textwidth]{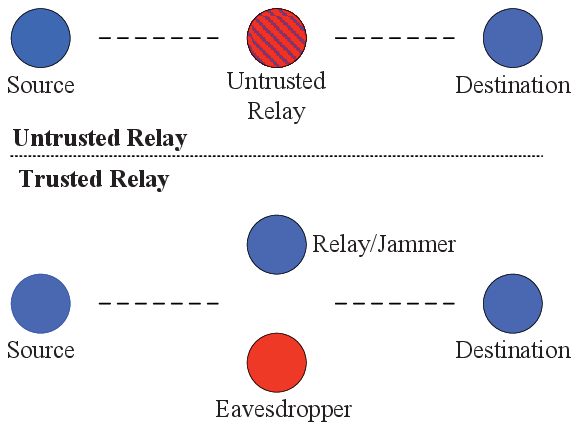}
\caption{Representations of trusted (distinct relay and eavesdropper) and untrusted (co-located relay and eavesdropper) relay networks.}
\label{fig_Trusted_UntrustedRelays}
\end{figure}

\subsection{Untrusted Relays}
As a pessimistic assumption, the relay itself can be considered to be an \emph{untrusted} user that acts both
as an eavesdropper and a helper, i.e., the eavesdropper is co-located with the relay node as shown in Fig.~\ref{fig_Trusted_UntrustedRelays}. The source desires to use the relay to communicate with the destination, but at the same time intends to shield the message from the relay. This type of model was first studied in \cite{Oohama_Capacity07} for the general relay channel. Coding problems associated with the relay-wiretap channel are studied under the assumption
that some of transmitted messages are confidential to the relay, and deterministic and stochastic rate regions are explicitly derived in \cite{He_Cooperation10,He_Two-Hop09,He_role08}, which showed that cooperation from
the untrusted relay is still essential for achieving a non-zero secrecy rate. In \cite{He_Cooperation10}, an achievable region of rate pairs $(R_1,R_e)$ was derived for the general untrusted relay channel.

Based on this region, the cooperation of an untrusted relay node was
found to be beneficial for a specific model where there is an
orthogonal link in the second hop. Cooperative relay broadcast channels
are discussed in \cite{Ekrem_Secrecy08}, where the users are untrusted
but not malicious. In such scenarios, users are willing to help each
other, but would not be allowed to decode each other's
message. Assuming a half-duplex amplify-and-forward protocol, another
effective countermeasure in this case is to have the destination jam
the relay while it is receiving data from the source. This intentional
interference can then be subtracted out by the destination from the
signal it receives via the relay.

In \cite{Jeong12}, the authors considered the joint source/relay beamforming design problem for secrecy
rate maximization, through a one-way untrusted MIMO relay. For the two-way untrusted
relay case, \cite{Ma13} proposes an iterative algorithm to solve for the joint beamformer
optimization problem, and \cite{HuangAsilomar13a} considers joint optimization for beamformer and untrusted relay node selection. In realistic fading channels, the secrecy outage probability (SOP) is
more meaningful compared with the ergodic secrecy rate, which is ill-defined under finite
delay constraints. Thus \cite{HuangTSP13} focuses on the secrecy outage probability of the AF relaying protocol, which is chosen
due to its increased security vis-\`{a}-vis decode-and-forward relaying and its lower complexity
compared to compress-and-forward approaches. As in Secs.~\ref{subsec:SISOWiretap} and \ref{sec:MIMOWiretap}, the SOP indicates the
fraction of fading realizations where a secrecy rate $R$ can be supported, and provides a
security metric when the source and destination have no CSI for the eavesdropper. The secrecy rate performance of untrusted relay selection was examined in \cite{Niu12}. In \cite{MukherjeeCommL13}, a constant BER of 0.5 is maintained at the untrusted relay by revealing to it only the real or imaginary components of the confidential $M$-ary symbols.

\subsection{Trusted Relays and Helpers}
Unlike the aforementioned case, in a \emph{trusted} relay
scenario the eavesdroppers and relays are separate network entities. Some of the most commonly encountered relay-based wiretap scenarios and corresponding solutions are depicted in Fig.~\ref{fig_mix}.
The relays can play various roles to counteract external eavesdroppers. They
may act purely as traditional relays while utilizing help from other
nodes to ensure security; they may also act as both relaying
components as well as cooperative jamming partners to enhance the
secure transmission; or they can assume the role of stand-alone \emph{helpers} to
facilitate the jamming of unintended receivers.

A typical model of a relay channel with an external eavesdropper was investigated by Lai \emph{et al.} in \cite{Lai_Relay--Eavesdropper08}, where
outer-bounds on the optimal rate-equivocation region are derived assuming a classical decode-and-forward protocol. The authors of \cite{Lai_Relay--Eavesdropper08} also propose a novel noise-forwarding strategy where the full-duplex relay sends dummy codewords independent of the secret message in order to confuse the eavesdropper. Such a strategy is also referred to as `deaf cooperation' in \cite{Bassily13,Bassily12}.

In \cite{Dong_Improving10,PetropoluWeber10}, several cooperative
schemes are proposed for a two-hop multiple-relay network, and the corresponding relay weights are derived to maximize the achievable secrecy rate, under the constraint that the link between the source and the relay is not protected from eavesdropping. The secrecy scaling laws in the limit of a large number of nodes for such a scenario are analyzed in \cite{Goeckel12}. The extension to a scenario with multiple eavesdroppers and maximum secrecy rate beamforming was pursued in \cite{MaSPL13}. It was shown in \cite{Mo12} that the decode-and-forward strategy is always outperformed by randomize-and-forward relaying (source and relay use different codebooks) in terms of secrecy outage probability.  \cite{Mo12} also discusses where to ideally place the relay.  In \cite{Ding12}, optimal precoding matrices based on artificial noise alignment are designed for a MIMO relay channel where the source, relay, and destination cooperatively jam an external eavesdropper, while robust relay beamforming was considered in \cite{Zhang13}. A combination of source GSVD precoding and relay SVD precoding was adopted in \cite{Jilani12} for the MIMO relay wiretap channel. A relay-assisted OFDMA downlink was considered in \cite{Schober11}, where the base station and relays jointly optimize the resource allocation for artificial noise versus data. Secrecy rate regions for a generalized relay network with parallel channels between all four terminals are derived in \cite{Vandendorpe12}. In \cite{Chockalingam13}, the set of relays is optimally divided into actual AF or DF relays and cooperative jammers, under an imperfect CSI assumption. Relay selection is another important issue when multiple relays are available; the optimal selection policy assuming the DF protocol was provided in \cite{ZouICC13} and shown to be superior to conventional max-min relay selection, while an opportunistic relay selection scheme was shown to have vanishing secrecy outage probability as the number of DF relays grew in \cite{Hanzo14}. A more general scenario was considered in \cite{ZouJSAC13} for AF and DF relays, with single and multiple relay selection schemes and corresponding diversity orders being presented. \cite{HuangAsilomar13b} considered utilizing a buffer-aided relay to enhance both transmission efficiency and security for two-hop relay networks.

\begin{figure}[ht]
\begin{center}
\includegraphics[width=0.87\linewidth]{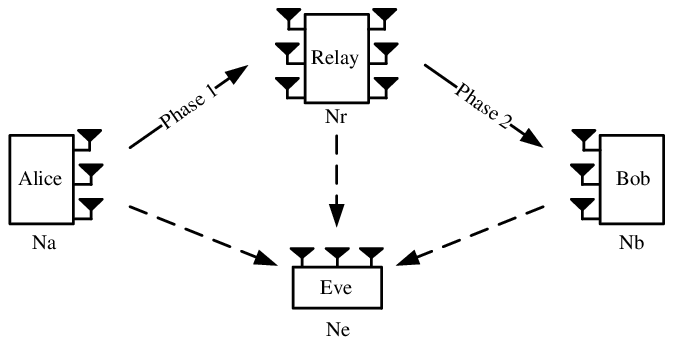}
\caption{\label{fig:sys_mod} Two-hop MIMO network with trusted relay and external eavesdropper.}
\end{center}
\end{figure}

Helpers serve as friendly jammers that do not have any information of their own to transmit, but instead cooperate with authorized nodes to degrade the signals intercepted by eavesdroppers. Namely, a helper can send a random codeword at a rate that ensures that it can be decoded and subtracted from the received signal by the intended receiver, but cannot be decoded by the eavesdropper.  Alternatively, a helper can transmit a jamming signal that interferes with the ability of the eavesdropper to intercept and decode the desired signal. For example, in a single-antenna wiretap channel with external helpers, an interesting approach is to split the transmission time
into two phases. In the first phase, the transmitter and the intended
receiver both transmit independent artificial noise signals to the
helper nodes. The helper nodes and the eavesdropper receive different
weighted versions of these two signals. In the second stage, the
helper nodes simply replay a weighted version of the received signal,
using a publicly available sequence of weights. At the same time, the
transmitter transmits its secret message, while also canceling the
artificial noise at the intended receiver \cite{GoelN08}.

 In \cite{Tang_Interference-assisted08}, a wiretap channel with an independent helping jammer was considered. The interferer can send a random codeword at a rate that ensures that it can be decoded and subtracted from the received signal by the intended receiver but cannot be decoded by the eavesdropper. The optimal helper power allocation over parallel OFDM subchannels is derived in \cite{Renna13ICC}. Optimal jamming weights and positions for helpers with mobility are presented in \cite{Petropulu13}. The effect of CSI feedback delay on relay and helper selection was quantified in \cite{Wu13}. In \cite{PeiTSP14}, a MISO scenario with constrained limited feedback of CSI from the receiver was considered, and an adaptive bit-allocation policy was proposed to optimally divide feedback bits between the transmitter and helper channels. The full MIMO scenario with artificial noise jamming by a single multi-antenna helper was analyzed in \cite{FakoorianTSP11}. The jamming strategy of a multi-antenna helper powered by energy harvesting instead of a regular battery was optimized in \cite{MukherjeeAsilomar12}.

 For the proposed coordinated cooperative jamming scheme for MIMO ad hoc networks in \cite{Wang_Cooperative09}, when one pair of nodes are communicating with each other, all the nodes surrounding the legitimate receiver cooperate to interfere with the eavesdropper by sending jamming signals. Orthogonal information subspaces and jamming subspaces are broadcast across the network, and artificial noise is chosen to lie in the publicized jamming subspace such that there will be no interference at the destination when an appropriate receive beamformer is used. An uncoordinated cooperative jamming strategy is also proposed for the case where the public jamming subspace is unavailable. In this case, the AN is simply the right singular vector of the main channel corresponding to the smallest singular value. Both schemes have been shown to efficiently increase the secrecy capacity, even if the eavesdropper has knowledge of the associated subspaces. The authors of \cite{HuangTSP12} considered a MISO channel with and without an external helper, and obtained robust beamforming/jamming solutions via numerical methods for imperfect CSI scenarios.

\begin{figure}[ht]
\begin{center}
\includegraphics[width=0.9\linewidth]{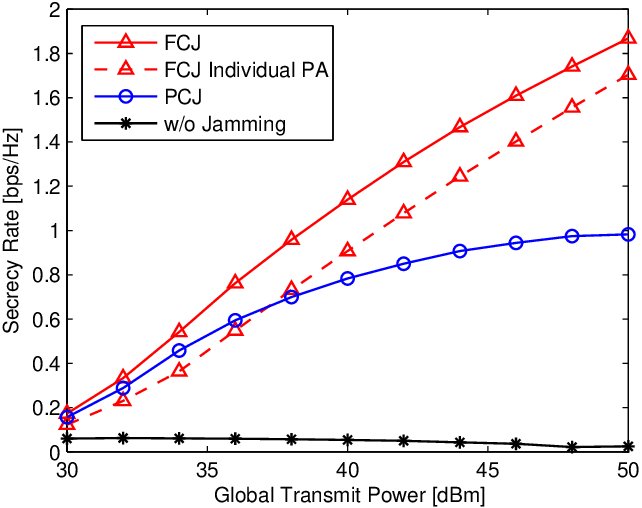}
\caption{\label{fig:Jack} Secrecy rate versus transmit power in two-hop channel with cooperative jamming, ECSIT unknown, four antennas at all nodes.}
\end{center}
\end{figure}

A more general case where cooperative jamming strategies guarantee secure communication in both hops without the need for external helpers was studied in \cite{Huang_Secure10}.
In these approaches, the normally inactive nodes in the relay network can be used as cooperative jamming sources to confuse the eavesdropper and provide better performance in terms of secrecy rate. In the proposed cooperative jamming strategies, the source and the destination nodes act as \emph{temporary helpers} to transmit jamming signals during transmission phases in which they are normally inactive. In \cite{PetropuluTIFS13}, the source transmits artificial noise along with data in the first hop, in addition to jamming by the destination. Jamming by the destination for the special case of a single-hop system was examined in \cite{GhoghoCOMML12,LiOttersten13}, which is feasible only when the destination has full-duplex capabilities, i.e., it can transmit and receive simultaneously on the same frequency with the aid of self-interference cancelation methods. Returning to \cite{Huang_Secure10}, two types of cooperative jamming schemes may be defined, \emph{full cooperative jamming} (FCJ) and \emph{partial cooperative jamming} (PCJ), depending on whether or not both the transmitter and the temporary helper transmit jamming signals at the same time. A comparison of these schemes is shown in Fig.~\ref{fig:Jack}.

In \cite{Tekin08b,Tekin08c}, a two-way wiretap channel was considered, in which both the source and receiver transmit information over the channel to each other in the presence of a wiretapper. Achievable rates for the two-way Gaussian channel are derived. In addition, a cooperative jamming scheme that utilizes the potential jammers was shown to be able to further increase the secrecy sum rate. \cite{He_role08a} showed that using feedback for encoding is essential in Gaussian full-duplex two-way wiretap
channels, while feedback can be ignored in the Gaussian half-duplex two-way relay channel with untrusted relays. More recently, secure transmission strategies are studied for the multi-antenna two-way relay channel with network coding in the presence of eavesdroppers \cite{Mukherjee_Securing10}-\cite{DebbahGC10}. By applying the analog network-coded relaying protocol, the end nodes exchange messages in two time slots. In this scenario, the eavesdropper has a significant advantage since it obtains two observations of the transmitted data compared to a single observation at each of the end nodes. As a countermeasure, in each of the two communication phases the transmitting nodes jam the eavesdropper, either by optimally using any available spatial degrees of freedom, or with the aid of external helpers.

\section{Wireless Secret Key Agreement}\label{sec:Key}

We recall that the original secure communication system studied by Shannon was based on secret-key encryption. Shannon's result that perfect secrecy required encryption with a random one-time pad cipher at least as long as the message was widely regarded as a pessimistic result, until it was reexamined in the context of noisy channels by Maurer \cite{Maurer93}.
In his seminal work, Maurer decried Wyner's degraded wiretap channel as being too unrealistic, and instead proposed a secret-key agreement protocol that could be implemented over a noiseless but authenticated and publicly observable two-way channel in the presence of a passive eavesdropper.

The key elements of Maurer's strategy are the \emph{information reconciliation} and \emph{privacy amplification} procedures. The information reconciliation phase is aimed at generating an identical random sequence between
Alice and Bob by exploiting a public discussion channel (sometimes split into a separate \emph{randomness sharing} step). The privacy amplification stage extracts a secret key from the identical random sequence
agreed to by two terminals in the preceding information reconciliation phase. In other words, after public discussion based on \emph{correlated randomness} in the first stage, privacy amplification reduces an initial piece of random nature into a smaller entity (e.g., by linear mapping and
universal hashing) which is known only by the legitimate
users, even if the eavesdropper has a less noisy channel in certain cases.

\begin{figure}[htp]
\centering
\includegraphics[width=0.9\linewidth]{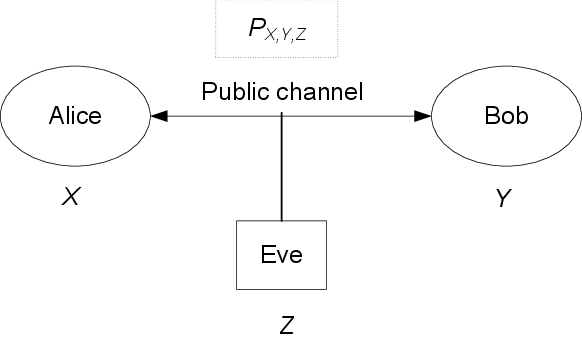}
\caption{Secret key agreement by $t$ rounds of public discussion between Alice and Bob. $X$ and $Y$ comprise the source of common randomness; Eve has access to $Z$ and joint distribution $P_{X,Y,Z}$.}
\label{fig_Maurer}
\end{figure}
More precisely, it was assumed that the transmitter, receiver and adversary have access to repeated independent realizations of random variables $X, Y,$ and $Z$, respectively, with some globally-known joint probability distribution $P_{X,Y,Z}$ as in Fig.~\ref{fig_Maurer}. The eavesdropper is completely ignorant of $X$ and $Y$. Alice and Bob undergo multiple rounds of two-way communication over the public channel, followed by generation of a shared key based on their individual information and observed messages.
The secret-key rate $S(X; Y||Z)$ between $X$ and $Y$ with respect to $Z$ is then defined as the maximal rate at which Alice and Bob can generate a secret
key over the noiseless public
channel in such a way that the adversary obtains information about this key only
at an arbitrarily small rate (cf. (\ref{eq:Korner1978})). The following upper and lower bounds on the secret key rate were presented in \cite{Maurer93}:
\begin{equation}
S\left( {X;Y|| Z} \right) \leq \min \left[ {I\left( {X;Y} \right),I\left( {X;Y|Z} \right)} \right],
\end{equation}
\[
S\left( {X;Y|| Z} \right) \geq \max \left[ {I\left( {X;Y} \right) - I\left( {X;Z} \right),I\left( {Y;X} \right) - I\left( {Y;Z} \right)} \right].
\]

Closely related results were offered in the concurrent work by Ahlswede and Csiz\'{a}r \cite{Ahlswede93}. Csisz\'{a}r and Narayan studied the augmentation of key-based secrecy capacity with the aid of a helper which supplies additional correlated information in \cite{Csizar00}, and obtained a single-letter characterization of the key-based secrecy capacities with an arbitrary number of terminals in \cite{Csiszar04}. Maurer and Wolf subsequently extended the secret-key sharing analysis of \cite{Maurer93} to account for the presence of an active eavesdropper in \cite{Maurer03I}-\cite{Maurer03III}, and showed that either a secret key can be generated at the same rate as in the passive-adversary case, or such secret-key agreement is infeasible. Refinements to their model that yield larger key rates are shown in \cite{Yakovlev08}. A two-user interference channel with a noiseless, shared feedback channel from the receivers and corresponding bounds on the secret-key capacity region are studied in \cite{Skoglund13}, while the multiple-access channel was examined in \cite{Salimi11}.

The next evolution in secret-key sharing was the exploitation of the common randomness inherent in reciprocal wireless communication channels.
Koorapaty \emph{et al.} relied on the independence of the channels between transmitter/receiver and transmitter/eavesdropper to use the phase of the fading coefficients as a secret key \cite{Hassan00}. Other techniques include key generation via
\begin{itemize}
\item discretizing extracted coefficients of the multipath components \cite{Shah07},
\item quantizing the channel phases  for a multitone communication system such that multiple independent phases are
used to generate longer keys \cite{Sayeed08},
\item directly quantizing the complex channel coefficients \cite{Shah06},
\item a purposely constructed random variable whose realizations are communicated between the legitimate
nodes, with secrecy achieved when the eavesdropper lacks channel state information \cite{Mclaughlin08},
\item exploiting the level crossing rates of the fading processes at the legitimate terminals \cite{Trappe10},
\item inducing more rapid fluctuations in channels from which keys are to be extracted via transmit array optimization \cite{Sasaoka05},
\item utilization of channel estimates as correlated random variables for information reconciliation \cite{Sasaoka10},
\item utilizing appropriately timed one-bit feedback available in practical networks due to Automatic Repeat reQuest (ARQ) protocols \cite{Abdallah10},
\item using unknown deterministic parameters such as wideband multipath channel parameters that are estimated by both Alice and Bob \cite{WinJSAC13}. This is a departure from the common randomness framework of Maurer, and a new notion of intrinsic information is defined accordingly to quantify achievable secret-key lengths.
\end{itemize}

Not surprisingly, multiple-antenna channels have attracted considerable attention for their capabilities of increasing common randomness at the legitimate users. The MIMO secret-key capacity for Gaussian inputs and system model identical to that of \eqref{eq:yHe} is \cite{Renna13}
 \begin{equation}\label{eq:MIMOseckeycap}
 \begin{split}
C_{SK} =& \mathop {\max }\limits_{{\mathbf{Q}}_x  \succeq 0} \log \det \left( {{\mathbf{I}} + {\mathbf{H}}_x {\mathbf{Q}}_x {\mathbf{H}}_x^H } \right) \\
       &{-}\: \log \det \left( {{\mathbf{I}} + {\mathbf{H}}_e {\mathbf{Q}}_x {\mathbf{H}}_e^H } \right).
\end{split}
\end{equation}
where $\mathbf{H}_x^H \mathbf{H}_x = \mathbf{H}_b^H \mathbf{H}_b + \mathbf{H}_e^H \mathbf{H}_e$ is an equivalent channel. Note the similarity to \eqref{eq:MIMOsecCap}, based on which a similar GSVD-based transmission scheme was adopted in \cite{Renna13}.
 Li and Ratazzi \cite{Ratazzi05} designed a randomized beamforming scheme based on knowledge of the main channel that makes blind channel estimation by the eavesdropper more difficult; the keyless secrecy rate of this method was examined in \cite{HuangCOMML13}. Chen and Jensen developed practical key generation protocols for MIMO systems with temporally and spatially correlated channel coefficients in \cite{Chen09,Jensen10}. Some of the first experimental measurement campaigns on secret key generation in reciprocal MIMO channels are presented in \cite{Patwari10,Sharma10}.

Previously discussed techniques for keyless security can be reutilized to enhance secret-key rates. The cooperative jamming method of \cite{Tekin08b} was used in \cite{Bloch11} as a precursor to secret key establishment over a two-way wiretap channel, and artificial noise was used to enhance secret key rates in a two-way relay network in \cite{Iwai11}. From the adversary's perspective, the optimality of Gaussian jamming against secret key establishment in two-way wireless channels was given in \cite{ZhouAsilomar13}. The role of a feedback channel in improving the secrecy rate of a wiretap channel has also been revisited in recent work. For a modulo-additive channel \cite{Lai08}, the authors showed that a noisy feedback channel that is observable by all parties can still be utilized to generate a secrecy rate equal to the main channel capacity, since the feedback from the (either full- or half-duplex) receiver acts as a private key that jams the eavesdropper. Ardestanizadeh \emph{et al.} \cite{Javidi09} considered a secure but rate-limited feedback channel, and proved that it is optimal for the receiver to feedback a random secret key that is independent of its received channel output symbols.

\section{Code Design for Secrecy}\label{sec:CodeDesign}
\subsection{Channel Coding}
Much like conventional networks, error correction codes play an integral role in building ``real-world" secure systems.
The McEliece cryptosystem \cite{McEliece78,Monico00} devised in 1978 can now be seen to be a bridge between channel coding-based physical layer security and classical cryptography. In this setup, the size-$(k \times n)$ generator matrix of a $(n,k)$ Goppa (linear) code capable of correcting $t$ errors is multiplied from the left and right by a randomly generated non-singular matrix and permutation matrix respectively, and the size-$(k \times n)$ product is made available as a public key. Messages sent to this entity are generated using the public key and then perturbed by a random vector of Hamming weight $t$. The ciphertext is decoded by multiplications with the inverses of the permutation and non-singular matrices interspersed with the code decoding algorithm.

Once the groundwork had been laid for the foundations of information-theoretic security [cf. Sec.~\ref{subsec:SISOWiretap}], several researchers turned their attention to the development of practical secrecy-preserving channel codes. Wyner \cite{Wyner75} and Csisz\`{a}r and K\"{o}rner \cite{Csiszar78} had used a stochastic coding argument to provide a non-constructive proof of the existence of channel codes that guarantee both robustness to transmission errors and a prescribed degree of data confidentiality as the block length tends to infinity.

In Wyner's stochastic encoding scheme, a mother codebook $C_0(n)$
of length n is randomly partitioned into ``secret bins" or subcodes
$\left\{C_1(n),C_2(n),\ldots,C_M(n)\right\}$. A message $w$ is associated
with a sub-code $C_w(n)$ and the transmitted codeword
is randomly selected within the sub-code. The
mother code $C_0(n)$ provides enough redundancy so that the
legitimate receiver can decode the message reliably, whereas
each sub-code is sufficiently large and, hence, introduces
enough randomness so that the eavesdropper's uncertainty
about the transmitted message can be guaranteed. However, the development of practical wiretap codes for general wiretap channels was not as rapid as that of classical error-correction codes in the two decades following Wyner's work.

Therefore, it was natural to turn to known capacity-achieving channel codes and examine their applications for secrecy \cite{HarrisonMag13}. In \cite{Thangaraj07}, Thangaraj \emph{et al.} advanced the idea of using graph-based codes such as low density parity check (LDPC) codes for binary erasure wiretap channels (noiseless main channel), and showed that both reliability and Wyner's weak secrecy criterion could be satisfied simultaneously. Bloch and coauthors \cite{Mclaughlin08} adopted LDPC codes and multi-level coding for the information reconciliation phase of a practical secret key agreement protocol. For Gaussian wiretap channels, appropriately punctured LDPC codes were employed with the relative bit error rate at the receiver and eavesdropper as a proxy security metric in \cite{Klinc09}, where the authors showed that a `security gap' was achievable. A turbo code-based scheme with the puncturing pattern determined by a pre-shared secret key was presented in \cite{ArefIEE}, while the achievability of high equivocation rates (cf. \eqref{eq:EquivocationRate}) with random puncturing was shown in \cite{Almeida13}.

Graph-based unstructured codes are not the only viable approach for wiretap coding. He and Yener \cite{He09StructCodes} showed that an arbitrarily large secrecy rate is achievable for Gaussian wiretap channels with an external helper using structured integer and nested lattice codes. Nested lattice codes were also deployed over the binary symmetric wiretap channel in \cite{Liu08}. Arora and Sang presented the notion of dialog codes wherein the receiver aids the transmitter by jamming the eavesdropper while still being able to recover the transmitted symbol \cite{Arora09}. If the receiver is half-duplex, then this can be achieved using a rate-1/2 code with memory where the receiver jams either of the code bits but is able to recover the message from the remaining bit, whereas the equivocation at the eavesdropper is unity. The recently proposed polar coding scheme has been shown to achieve the secrecy capacity for binary symmetric and deterministic wiretap channels \cite{Vardy,FakoorianAsilomar13}. Polar coding was subsequently extended to secret-key generation in \cite{Abbe13}, and shown to be secret-key capacity-achieving for a binary symmetric channel.

\subsection{Distributed Storage Coding}
\begin{figure}[htp]
\centering
\includegraphics[width=\linewidth]{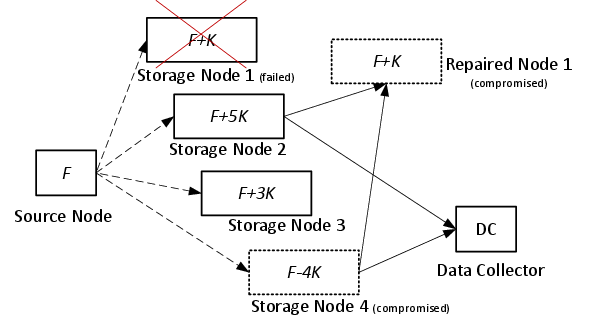}
\caption{Security problem in distributed storage network with an eavesdropper that can observe data in compromised nodes.}
\label{fig_DistribStorage}
\end{figure}

A recent avenue for coding theory research is the design of resilient codes for distributed data and cloud storage systems. The essence of such systems is that chunks of data files are scattered across various storage nodes, and it is desired that an end-user or data collector be able to accurately reconstruct the original files by retrieving data from a subset of $k$ such storage nodes. However, the storage nodes are assumed to be unreliable and prone to failure (equivalent to data erasures), and thus fault-tolerance to such failures under bandwidth constraints is the primary code design criterion. These considerations lead to the introduction of a new class of `regenerating codes' which are efficient with
respect to both storage space utilization and the amount of data
downloaded for repair (termed repair-bandwidth) \cite{Dimakis10}. In addition to reliability, it is also critical to protect data from being reconstructed by eavesdroppers. A passive eavesdropper that can access the data on up to $\ell$ storage nodes is denoted a Type-I adversary in \cite{Tandon13}, and as a Type-II adversary if it can also observe the repair data of $\ell$ nodes. A typical security scenario is shown in Fig.~\ref{fig_DistribStorage}, where $\ell=2$ out of $k=4$ storage nodes have been compromised by an eavesdropper that seeks to reconstruct the original file $F$.

Pawar \emph{et al}. studied the problem of securing distributed storage systems against eavesdroppers and malicious adversaries in \cite{Ramchandran12}, and defined the secrecy capacity $C_s\left( {\alpha ,\gamma } \right)$ as the maximum amount of data that can be stored in the system such
that the data can be reconstructed reliably while remaining perfectly secret from Eve, for all possible data
collectors and eavesdroppers. Their upper bound on the system secrecy capacity for a Type-I adversary turned out to be
\[{C_s}\left( {\alpha ,\gamma } \right) \leq \sum\limits_{i = \ell + 1}^k {\min \left\{ {\left( {d - i + 1} \right)\frac{\gamma }{d},\alpha } \right\}}\]
where $\alpha$ is the storage capacity in symbols of each of the $k$ total nodes, and $\gamma$ is the total amount of data downloaded for repair by the replacement node from $d$ unaffected storage nodes. The bound verifies the intuition that only the $k-l$ non-compromised nodes can yield secure and reliable information to the data collector.
Shah \emph{et al.} constructed secure exact repair codes based on the product-matrix framework in \cite{Rashmi11}, which ensures that
the information contained in the symbols downloaded by the replacement node is independent of the identities of the helper
nodes. Dikaliotis \emph{et al}. studied the security of distributed storage systems in the presence of a trusted verifier \cite{Dimakis10}. The maximum file size that can be stored securely was determined for systems in which all the available nodes help with repair in \cite{Goparaju13}. The single node repair setting was generalized to multiple node failures and secrecy capacity bounds provided for the same in \cite{Koyluoglu12}. The characterization of the secure storage-vs-exact-repair-bandwidth tradeoff region under both Type-I and Type-II attacks was given in \cite{Tandon13}. For a heterogeneous system
with nodes having different storage capacities and different repair bandwidths, lower and upper bounds on the system capacity were given in \cite{Ernvall13}.

\subsection{Network Coding}
While the emerging area of network coding is not directly related to traditional channel coding design \emph{per se}, we briefly mention physical layer security issues encountered in this field. Network coding is a paradigm for multi-hop wireline and wireless networks that allows intermediate nodes to `mix' packets or signals received from multiple paths, with the objective of improving throughput \cite{Yeung}. Therefore, such networks are vulnerable to eavesdropping, akin to other networks discussed thus far in this work.

The secure network coding problem was introduced in \cite{Cai02} for multicast wireline networks where each link has equal capacity, and a wiretapper can observe an unknown set of up to $k$ network links. For this scenario, the secrecy capacity is given by the cut-set bound, and is achieved by injecting $k$ random keys at the source which are decoded at the sink along with the message \cite{Cai02,Stein04}. Silva and Kschischang \cite{Kschischang08} among others have drawn connections between the multicast problem and the type-II wiretap channel studied by Ozarow and Wyner, as described in Section~\ref{subsec:SISOWiretap}. Eavesdropping countermeasures for wireless network coding systems are described in \cite{Mukherjee_Securing10,Zhang09}, among others. In \cite{Liu11}, a distributed version of the randomized transmission scheme of \cite{Ratazzi05} was adopted for a cooperative network coding system with external eavesdroppers, with bit error rate as the performance metric.

\section{Related Topics}\label{sec:RelatedTopics}
\subsection{Game Theory and Security}
The interactions between various agents (transmitters, receivers, helpers, and attackers) in multiuser wireless networks are accurately captured by inter-disciplinary analyses based on game theory and microeconomics, and this holds true for problems of secrecy as well. The central tenet of game theory is to model agents or players as rational entities whose sole focus is to maximize their individual gains or payoff functions. A non-cooperative game model assumes agents eschew coordination with one another (e.g., in a 2-player zero-sum game the payoffs add up to zero), while in a cooperative game players may choose to cooperate to achieve some mutual benefit (e.g., players may offer monetary payments via an auction, or form a coalition). Stable outcomes from which no player has an incentive to deviate are known as Nash Equilibria.

A zero-sum game between a multi-channel transmitter and an adversarial nature in the presence of an eavesdropper was treated in \cite{Garnaev09}, with the difference of Alice and Eve's SINR as the payoff. Utilizing secrecy rate as the payoff in a game-theoretic formulation is a relatively new concept. Yuksel, Liu, and Erkip studied a SISO wiretap network with an adversarial jammer helping the eavesdropper as a zero-sum game, and presented the Nash Equilibrium input and jammer cumulative distribution functions \cite{Erkip09}. In \cite{MukherjeeICC,MukherjeeTSP13}, the authors considered a MIMO wiretap channel with an active eavesdropper that can either listen or jam, and pose its interactions with the transmitter as a zero-sum game with the MIMO secrecy rate as the payoff function. The SISO one-sided interference channel was studied in \cite{UlukusISIT11}, and the corresponding Nash equilibrium secrecy rate region was derived.  A zero-sum power allocation game between a multi-channel transmitter and a hostile jammer that is distinct from the eavesdropper was formulated in \cite{Ara12}, with the secrecy rate as the payoff function.

Cooperative game theory was applied in \cite{Saad09} to demonstrate the improvement in secrecy capacity of an ad hoc network, when users form coalitions to null the signals overheard by eavesdroppers via collaborative beamforming. For a hierarchical multi-hop system with different potential paths to the base station, a distributed tree formation game was postulated in \cite{Maham12}. Han \emph{et al.} \cite{Debbah09} developed a two-stage Stackelberg game where a transmitter `pays' a number of external helpers to jam an eavesdropper, and computed the corresponding equilibrium prices and convergence properties. The same authors examined a similar scenario in \cite{Marina09}, where an auction game was used instead to model the transactions between transmitters and helping jammers. Anand and Chandramouli studied an $M$-user non-cooperative power control game with secrecy considerations in \cite{Anand}, and applied pricing functions to improve the energy efficiency and sum secrecy capacity of the network. For the 2-user IC with confidential messages, we have discussed in Sec.~\ref{sec:IC} how Kalai-Smorodinsky bargaining solutions and zero-sum games are adopted to allow the transmitters to find an operating point that balances network performance and fairness \cite{IFCAlic10,FakoorianTSP13MISO}. In \cite{Hong11}, game theory is used by multiple eavesdroppers to decide whether to collude or not in a MISO wiretap channel.

\subsection{Cognitive Radio and Sensor Networks}
As a promising technique to alleviate spectrum scarcity, cognitive radio (CR) \cite{Mitola_Cognitive99} is capable of dynamically sensing and locating unused spectrum segments in a target spectrum pool and communicating using the unused spectrum segments in ways that cause no harmful interference to the primary users of the spectrum. Due to the vulnerability of CR physical layer spectrum sensing, research attention on physical layer security issues, though limited, has emerged recently. In \cite{Clancy_Security08,Askoxylakis13}, several classes of physical layer attacks for
dynamic spectrum access and adaptive radio scenarios are described, and corresponding techniques to mitigate these attacks are proposed.  Denial-of-service vulnerabilities from the perspectives of the network architecture employed, the spectrum access technique used and the spectrum awareness model assumed, are examined in \cite{Brown_Potential07} and possible remedies are provided. Achievable secrecy rates in CR networks with external eavesdroppers have been studied in \cite{Chandramouli08,Pei10}.

Network spectral efficiencies can be further improved if the cooperative jamming signals are data signals instead of indiscriminate artificial noise. An elegant example of such a system is a CR network where the primary user wishes to conceal its message from an external eavesdropper \cite{Yongle11,Stanojev13}. Here, the role of helpers is played by secondary or unlicensed users that seek to opportunistically transmit their data in the frequency band occupied by the primary user. Since the eavesdropper is interested only in the primary message, the secondary user signals act as jamming signals at the eavesdropper (as well as the primary receiver). The primary signal in turn is perceived as interference at the secondary receivers. Therefore, it is critical to judiciously select the primary and secondary signal powers in tandem so as to maximize the joint rate region of  the cooperating users. In \cite{Yongle11,Stanojev13}, this was achieved via a Stackelberg power-control game formulation for the primary-secondary interactions, where the primary user allows secondary transmissions only if its secrecy rate is improved by doing so. In the multi-channel scenario of \cite{Basar13}, the primary users are oblivious to the presence of CRs, while a game-theoretic formulation was constructed for optimal channel selection by the CRs and external eavesdroppers.

While not directly related to information security, a so-called \emph{primary user emulation} (PUE) threat to spectrum sensing was identified in \cite{Chen_Defense08}. In PUE, a malicious node mimics the signal characteristics of licensed users in order to mislead cognitive radios into vacating the spectrum. As a countermeasure, \cite{Chen_Defense08} proposed a transmitter verification scheme to verify whether a given signal is that of an incumbent transmitter by estimating its location and observing
its signal characteristics. Another major physical-layer vulnerability in cooperative spectrum-sensing CR systems is the deliberate feedback of false sensing information. In \cite{Wang_PHY-layer10}, this problem is solved by designing fusion center (FC) counting rules so as to minimize
the overall false alarm probability. Details of security challenges peculiar to cognitive radio networks can be found in \cite{Askoxylakis13}.

Wireless sensor networks and corresponding distributed estimation algorithms have been at the forefront of signal processing research in the past decade. The downlink and uplink phases of communication between the sensors and the FC are inherently vulnerable to eavesdropping. Li, Chen, and Ratazzi \cite{RatazziSensor05} tackled downlink secrecy when the FC has multiple antennas by deliberately inducing rapid time-varying fluctuations in the eavesdropper's channel. \cite{Ives05} proposed the use of artificial noise-like schemes on the uplink to `confuse' eavesdroppers about the aggregate sensor observations sent to the FC. In \cite{Naraghi12}, the sensor observations are randomly mapped to a set of discrete quantization levels, with the corresponding mapping probabilities known only to the intended FC and not the eavesdropper. The optimal mapping probabilities and FC decision rule that jointly minimize its error probability subject to a constraint on the eavesdropper error probability are then derived. Marano \emph{et al.} \cite{Marano09} examined optimal sensor censoring strategies in an energy-constrained sensor network infiltrated by an eavesdropper. Kundur \emph{et al.} examined cross-layer secrecy-preserving design methodologies for multimedia sensor networks in \cite{Kundur08}.

\subsection{Secrecy in Large-Scale Networks}\label{sec:StochGeom}
Unlike point-to-point scenarios, the communication between nodes in large-scale networks
strongly depends on the location and the interactions between nodes. In an early work on eavesdropping in cellular CDMA networks with multi-user detection capabilities, the outage probability of the eavesdropper signal-to-interference ratio was adopted as the performance metric \cite{Verdu98}.
Based on the assumption that legitimate nodes and eavesdroppers are distributed randomly in
space, studies on secure communications for large-scale wireless networks have been
carried out recently, utilizing tools from stochastic geometry and graph theory. Analyses based on stochastic geometry typically assume a spatial point process model (e.g., Poisson) for the locations of transmitters and receivers.

Secrecy communication graphs describing secure connectivity over a large-scale network with eavesdroppers present were
investigated in \cite{Haenggi08}-\cite{Pinto12}. In particular, the statistical characterizations of in-degree and out-degree
under the security constraints were considered by Haenggi \cite{Haenggi08}, Pinto \emph{et al}. \cite{Pinto10} and
Goel \emph{et al}. \cite{Goel10}. By using tools from percolation theory, the existence of a secrecy graph
was analyzed in \cite{Haenggi08}, \cite{Goel10}. The results in \cite{Pinto12} showed improvements in secure connectivity
by introducing directional antenna elements and eigen-beamforming. In order to derive the
network throughput, these works on connectivity were further extended to incorporate secrecy capacity
analysis. Specifically, the maximum achievable secrecy rate under the worst-case scenario
with colluding eavesdroppers was given in \cite{Koksal12}. Scaling laws for secrecy capacity in large
networks have been investigated in \cite{Ying09}-\cite{Ganti11}. Focusing on the transmission capacity of secure
communications (defined as the number of successful transmissions taking place in the network per unit area, subject to a constraint
on secrecy outage probability), the throughput cost of achieving a certain level of security in an interference-limited
network was analyzed in \cite{Zhou13}, and the impact of uncertainties in node positions and CSI was examined in \cite{Abreu13}. The probability of secure connectivity was given in \cite{Andrews11} for multi-antenna nodes, and in \cite{Cai13} for a scenario with randomize-and-forward relays and a PPP for eavesdropper locations. A hierarchical multi-level sensor network was considered in \cite{Rabbachin13}, which introduced the concept of distributed network secrecy throughput to quantify inter-level network secrecy of all levels, i.e., data transmitted without
collision are received successfully without being successfully eavesdropped in all levels.

\subsection{Physical Layer Authentication}
The most critical aspect of physical layer security is to ensure that confidential messages are decoded only by their intended receivers. A less well-studied but necessary component is that of message authentication, namely, to enable the receiver of a message
to detect whether it was forged or illegitimately modified by someone other than its claimed source. Much like secure encoding, these operations are usually performed at the network and higher layers, with recent interest in devising physical layer counterparts.

In \cite{UMMaurer2000} an information-theoretic analysis of authentication
is provided assuming both that the legitimate transmitter
and receiver share a secret key and that transmission among all
parties (including the attacker) are noiseless. In \cite{Martinian05} both legitimate
distortions of the message and joint typicality decoding
are introduced in this framework. The impact of
both noise and errors in the channel was taken into account
for the first time in \cite{Poor_auth2009}. There, information theoretic bounds
on the probability of a successful attack were derived for an
arbitrarily low false alarm rate and infinitely long codewords.

Current attempts at using physical layer characteristics as
authentication keys for the message source follow various
approaches. One possibility is to assume a pre-shared secret
key hidden in the modulation scheme, which is detected
by the receiver \cite{Sadler08,Baras09}. In other keyless transmitter-based
methods (referred to as wireless fingerprinting), device-specific non-ideal
transmission parameters are extracted from the received
signal. They are identified as characteristics of the claimed
source and then compared with those from previous authenticated
messages \cite{Daniels05}. Channel-based authentication algorithms
compare the channel response estimated from the current
message with that estimated from the previous transmissions
by the ostensibly verified source, in effect authenticating the position
of the transmitter rather than its identity. In order to reliably
distinguish channels from different locations, some source
of diversity must be exploited, either in the spatial domain by
measurements of the received power levels at many receivers
\cite{Cheriton06}-\cite{Martin07} or in the frequency domain via wideband channel
estimates \cite{Xiao07}-\cite{Trappe2010}. Instead of explicitly using channel responses for authentication, Tugnait \cite{TugnaitJSAC13} distinguishes between message sources based on their power spectral densities. A summary of a wide range of possible methods is available in \cite{Mohapatra10}.

For the case of a multi-antenna channel, \cite{Tomasin12} considers an approach where the test is performed in two phases. In the first phase,
the receiver gets an authenticated noisy estimate $x$ of the channel with respect to the legitimate
transmitter. In the second phase, upon reception of a message, the receiver gets a new estimate
$u$ of the channel and compares it with $x$. A hypothesis test is subsequently performed to determine whether $u$ is an estimate of the
legitimate channel or the channel forged by an eavesdropper.

\section{CONCLUSIONS AND DIRECTIONS}\label{sec:concl}
This paper has provided a comprehensive survey of the field of physical layer security in wireless networks based on information-theoretic principles. We commenced with an overview of the foundations dating back to the pioneering work of Shannon, Wyner, and Maurer on information-theoretic security. We then described the evolution of secure transmission strategies from point-to-point channels to multiple-antenna systems, followed by generalizations to larger multiuser networks. We also reviewed secret-key establishment protocols based on physical layer mechanisms, along with an overview of practical secrecy-preserving code design and inter-disciplinary approaches for security. The associated problem of physical layer message authentication is also introduced. Broadly speaking, it was observed that physical layer security is achieved by either exploiting the independence of wireless channels and background noise conditions observed by different nodes, or by judiciously directing interference (exogenous or intentional) towards unintended receivers.

The scope for future work in this field is extensive and only a few select directions are discussed next. As an example, the application of physical layer security techniques to commercially deployed wireless systems is largely unexplored. The majority of the techniques discussed in this survey, such as artificial noise for eavesdropper jamming and CSI-based precoding to optimize secrecy rates, are agnostic to the underlying air interface (time/code/orthogonal frequency-division multiple access). For example, an OFDMA-based base station may choose to transmit artificial noise along with data symbols in certain subcarriers as long as spectral emission masks are not violated. A CDMA transmitter may do the same after spreading the data with a pseudo-noise sequence. Furthermore, in 3GPP LTE, the introduction of Demodulation Reference Symbols (DMRS) has enabled the use of arbitrary MIMO precoders by the base station, therefore the secure GSVD precoder of \cite{Khisti10} or its variants can be implemented without change in the current LTE standard. The secret-key generation scheme in \cite{ChengJSAC13} that makes use of LTE precoding matrix indicator (PMI) feedback is therefore a starting point for this direction. Arbitrary MIMO precoding is also allowed in IEEE 802.11ac and other forthcoming WLAN standards. The introduction of relay nodes, machine-type communications, and device-to-device communications in LTE raise new security challenges \cite{Cao14}, and conceivably heighten the need for combining physical layer security with existing key-based ciphers.

Indeed, since physical layer security issues arise in multiuser systems of any kind, it is expected that new network scenarios and corresponding security schemes will continue to be developed. For instance, massive MIMO systems, overlay cognitive radio networks, smart grid systems \cite{Gerla12}, networks with simultaneous wireless information and power transfer \cite{ZhangSWIPT13}, and heterogeneous networks \cite{Chen13} are untapped case studies from a secrecy perspective, to name just a few. Holistic approaches spanning the application and physical layer, in addition to exploitation of reconfigurable antennas \cite{Malyuskin12}, are expected to become more prominent. It is evident that the use cases of physical layer security extend well beyond cellular systems as seen in this survey.

Another untapped area is cross-layer analysis of secrecy combined with considerations of data queueing delay and rate control. In conventional network control problems, data packets that need to be served arrive in a queue(s) following some stochastic process, and the system is considered stable if the queue lengths are confined to some finite length. Initial steps to incorporate secrecy constraints into such problems were taken in \cite{YingTIFS11} for a broadcast channel with confidential messages, where a secrecy throughput-optimal scheduling scheme was provided under a network utility maximization framework. More recently, for a single-user scenario the authors of \cite{Shroff13} maximized the long-term data
admission rate, subject to the stability of the data queue as
well as a bound on the rate of secrecy outage. Evidently, many additional network scenarios await further analysis.

Finally, a deeper understanding of the interplay between physical layer security and classic cryptographic security is another rich but unexplored resource for further study \cite{Vardy10,Goldsmith13}. Also of current interest are secure transmission schemes where the confidential message also remains covert, i.e., potential eavesdroppers are uncertain if transmissions are on-going \cite{Jaggi13,Kramer13}.



\end{document}